\documentclass{osa-article}
\usepackage{amsmath}
\usepackage{multirow}
\usepackage{xcolor}
\usepackage{algorithm}
\usepackage{algorithmic}
\journal{osac}

\articletype{Research Article}

\newcommand{\figref}[1]{Fig.~\ref{#1}}
\newcommand{\eqnref}[1]{Eq.~(\ref{#1})}
\newcommand{\tabref}[1]{Tab.~\ref{#1}}

\begin{document}

\title{Measuring the cophasing state of a segmented mirror with a wavelength sweep and a Zernike phase contrast sensor.}

\author{Anne-Laure Cheffot,\authormark{1,2,*} Arthur Vigan,\authormark{1},  Samuel Leveque\authormark{2}, and Emmanuel Hugot\authormark{1}}

\address{\authormark{1}Aix Marseille Univ, CNRS, CNES, LAM, Marseille, France\\
\authormark{2}European Southern Observatory, Karl-Schwarzschild-Strasse 2, 85748 Garching, Germany}

\email{\authormark{*}acheffot@eso.org} 



\begin{abstract}
The demand for higher resolution telescopes leads to segmented primary mirrors which need to be phased for operation. A phasing sensor applying a wavelength sweep technique provides a large capture range without modulating the position of individual mirror segments. This technique offers the potential to monitor the phasing state of a segmented telescope in parallel to the science observations. 
We evaluate the performance of the wavelength sweep technique using a Zernike phase contrast sensor for coarse phasing. Tests results on a dedicated bench show 112 nm rms precision. With the help of a simulation, we explain a known error of the method and we suggest ways for improvements. 
\end{abstract}

\section{Introduction}\label{sec:intro}
The next generation of large optical and infrared telescopes \cite{ELT:18,ELT:WS,TMT:WS,GMT:WS,JWST:WS} rely on segmented primary mirrors, based on the heritage of the Keck\cite{Keck:91} and GTC\cite{GTC:04,GTCPhasing:ukn} telescopes. 
The relative height (piston) and orientation (tilt) between neighbouring mirrors segments need to be controlled to within a fraction of the wavelength of observation \cite{Bely:03}.
In the particular  case of the ELT, each of the 798 segments will be  equipped with six edge sensors and three position actuators. 
Their role is to maintain a piston and tilt set-point defined during periodical on-sky phasing runs with the "Phasing and Diagnostic Station" \cite{LewisSPIE:18}. 
This station will include a suite of different visible and near infrared sensors, performing guiding, high and low order wavefront sensing, as well as phasing tasks. 
Phasing runs compete with science time and are foreseen to occur typically every two weeks.

Several types of phasing sensors have been validated on sky using 10\,m-class telescopes, such as the Shack-Hartmann \cite{Chanan:98,Gonte:11,APESPIE:08}, the Pyramid \cite{APESPIE:08} and the Zernike phasing sensors \cite{Surdej:10,APESPIE:08,Dohlen:03}. 
These sensors provide comparable performances, reaching about 20\,nm rms piston error (wavefront) in closed-loop for a star as faint as 18th magnitude, reaching a piston error as low as 6\,nm rms for bright stars. 
Other types of sensors such as ELASTIC \cite{VievardSPIE:18} or the self-coherent camera \cite{Janin:16} have been tested and validated. ELASTIC has demonstrated that it could align and phase a seven-segments mirror with a residual wavefront error of around 50\,nm RMS. The Self-Coherent camera, on the other hand is derived from a four quadrant coronograph and in simulation demonstrates its capacity to phase a segmented mirror with less than 5\,nm residual RMS wavefront. 
Except ELASTIC, most phasing sensors' capture ranges are limited to $\pm$0.5 the sensing wavelength.
The ELT baseline wavefront control strategy foresees the implementation of a Shack-Hartmann phasing sensor, assisted by a Zernike phase contrast sensor for accurate pupil registration.\cite{BonnetSPIE:18} 

A freshly integrated segment will be mechanically aligned, with tens of micrometers precision. 
Most phasing sensors need an additional method to extend their range to tens of micrometers.
Various methods have been implemented for extending the capture range beyond several microns. 
These include implementing multicolour schemes \cite{Chanan:00,Vigan:11} and measuring light coherence by modulating the segments \cite{Bonnet:14,Chanan:98}. 
However, these methods rely on modulating the segments' piston and therefore cannot be used in parallel with science observations.  
This can be avoided by using a wavelength sweep technique as demonstrated by \cite{Bonaglia:10} in combination with a pyramid phasing sensor. 
The implementation is based on the commercial availability of a Liquid Crystal Tunable Filter (LCTF \cite{lctfThor:ukn}). 
This device allows the rapid setting of the wavelength between 650\,nm and 1100\,nm with a resolution of 1\,nm and a bandwidth of 10\,nm. 
A capture range of 15\,$\mu$m has been validated on-sky with a precision better than 0.25\,$\mu$m \cite{Bonaglia:10}.

The wavelength sweep technique has the potential to monitor the phasing state of a large segmented mirror telescope with a large capture range, without affecting science observations. 
This may help in identifying phasing performance degradation between periodic phasing runs, which may be difficult to detect through indirect failure detection schemes. 
In addition, it could help with pre-phasing freshly re-coated segments that will be installed on a daily basis. 
In this particular context, the Zernike phase contrast sensor represents a good sensor candidate because of its intrinsic simplicity. 
Its signal has the same dependence as the pyramid sensor's to the relative height between two segments. 
For example it could be integrated at the intermediate focus of one of the ELT's guide probes \cite{LewisSPIE:18} should they be upgraded in the future.

In section 2 we present the working principle of the wavelength sweep technique and how it can be applied to the Zernike phase contrast sensor. 
We describe the expected piston measuring range and introduce strategies for minimizing the impact of two known sources of piston measurement errors: a sign error and a random error, appearing for small pistons values.
In section 3 we show experimental results obtained on an upgraded version of ESO's APE test bench\cite{Gonte:04}. 
A comparison with simulations is given in section 4, before concluding and discussing next steps for validation.

For clarity purposes, the word piston will designate a segment reconstructed height with respect to a reference, the acronym optical path difference ($OPD$) the relative height between two neighbouring segments, and the term border will refer to the separation between two adjacent segments where the $OPD$ was measured.

\section{Wavelength sweep using a Zernike phase contrast: working principle}\label{sec:method}

\subsection{Wavelength sweep technique}\label{subsec:WST}

Previous studies \cite{PinnaSPIE:06,Bonaglia:10} report the use of the Wavelength Sweep Technique. 
The working principle is as follows: in most interferometer systems, the intensity of the fringes depends on the path difference between two surfaces and on the wavelength used. 
They are linked by the following mathematical expression:
\begin{equation}
    I = A \sin{\left( \frac{2 \pi}{\lambda} OPD +\phi \right)}+B = A \sin{\left( k OPD +\phi \right)}+B
    \label{eqn:WST}
\end{equation}
Where $A$ and $B$ are fit variables of respectively the sinusoidal amplitude and offset and $\lambda$ is the wavelength. 
$\frac{2 \pi}{\lambda}$ can also be denoted as $k$, the wavenumber. 
$OPD$ is the relative height between the two neighbouring segments and $\phi$ a free parameter for the phase delay. 

\begin{figure}[h!]
\centering
\includegraphics[width=7cm]{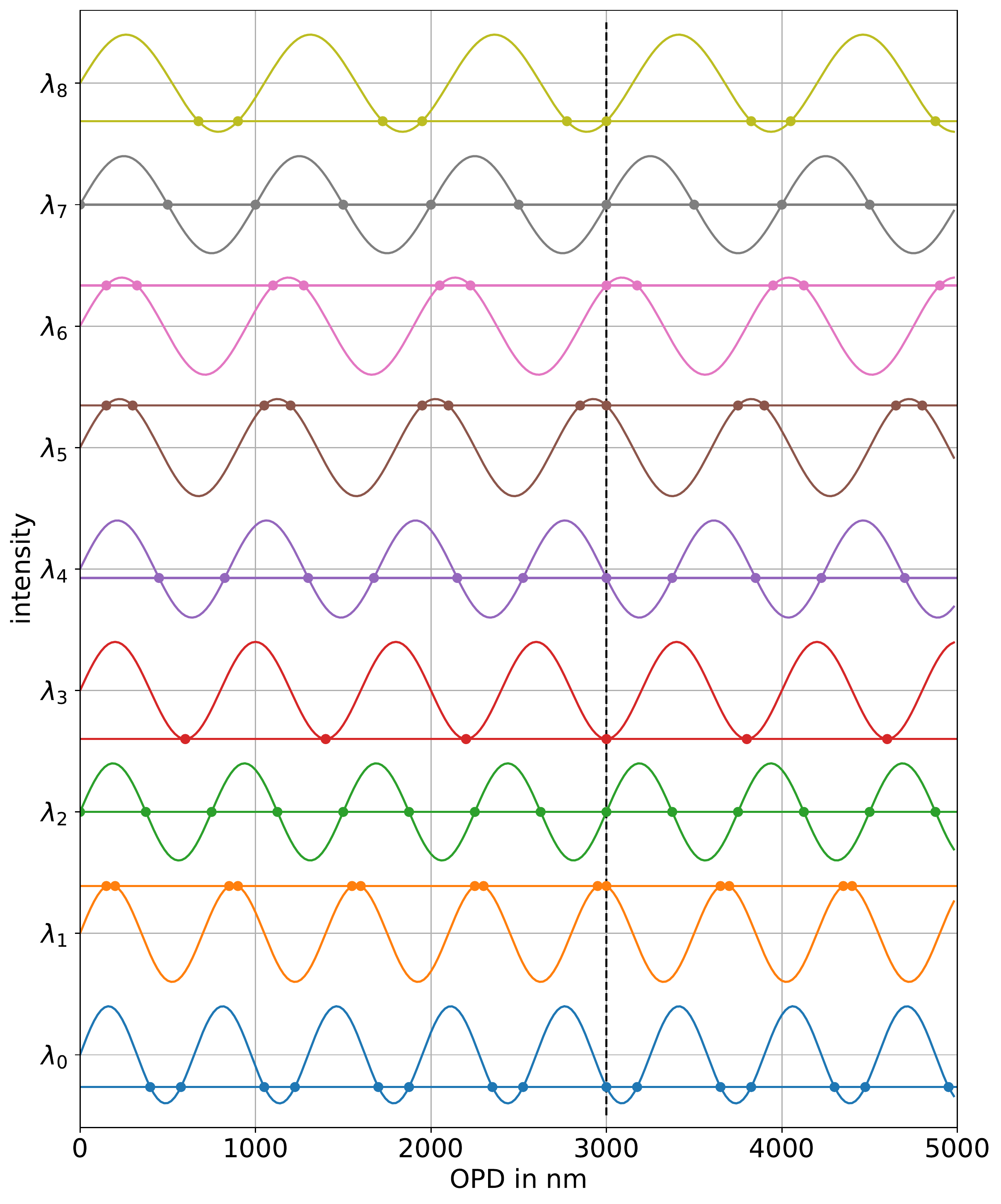}
\caption{Illustration of the excess fraction in an ideal case (no noise). Each colour represents a wavelength, and they have been vertically offset for readability. The wavelengths are sorted such that $\lambda_{0}<\lambda_{1} ...<\lambda_{8}$}. The sinusoid represents the variation of the intensity with the $OPD$ in x-axis. A measurement has been made at each wavelength. Since the signal has a sinusoidal dependency, the signal registered could have been produced by any of the $OPD$ with a round marker. The only solution to the problem is the one for which all wavelengths' likely solutions line-up as indicated by the black dashed line. This perfect alignment is only possible because there is no noise in this example.
\label{fig:WST1}
\end{figure}

Classically, a multi-wavelength method is analysed by the excess fraction method or derived methods (see \cite{Michelson:1895}). 
This approach is illustrated in \figref{fig:WST1}.

Each coloured sinusoid represents how the intensity of a fringe would vary with the $OPD$ indicated on the x-axis. 
For a given measurement, each wavelength creates a different intensity. 
This intensity could have been generated by any of the $OPD$s at which there is a round marker on \figref{fig:WST1}. 
One needs to look at all the wavelengths to identify which $OPD$ creates the measured intensities: it is the one for which all wavelengths and their corresponding intensities give the same $OPD$, as represented by the vertical dashed line in \figref{fig:WST1}. 

\begin{figure}[h!]
\centering
\includegraphics[width=7cm]{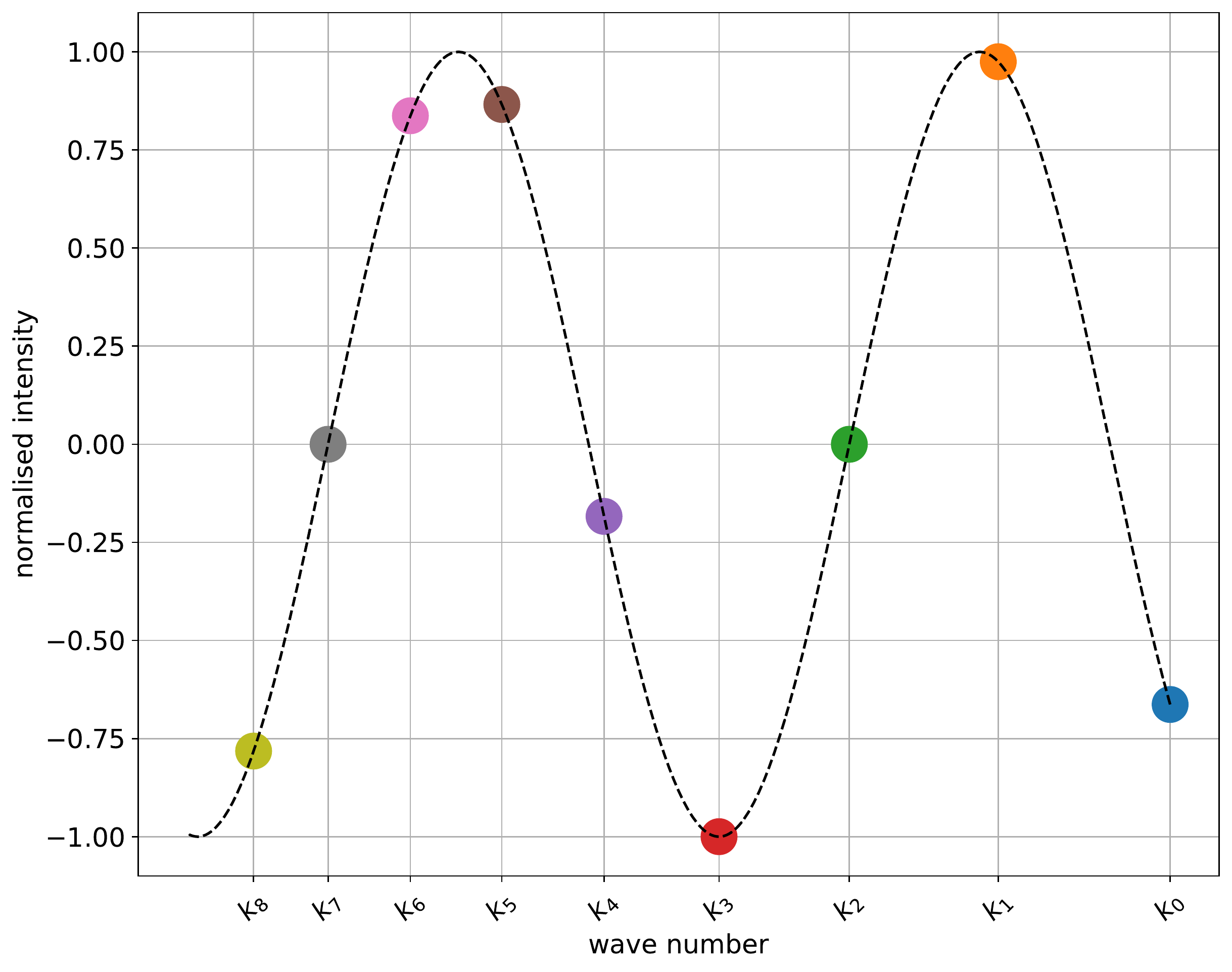}
\caption{Illustration of the wavelength sweep method as a function of wavenumber. The intensity variation of each wavelength is displayed by the round markers. The intensity variation evolves according to \eqnref{eqn:WST} shown by the black dashed line}
\label{fig:WST2}
\end{figure}

Another way of displaying the intensity measurements is illustrated in \figref{fig:WST2}.
The markers represent the recorded intensity for the same wavelengths as before, as a function of their wavenumbers. 
In this representation, the sinusoid's frequency is the $OPD$.
Because of the redundant nature of the sinusoidal fit, the search domain of the fitting needs to be restrained.
It is limited to only positive values for A and $OPD$ of \eqnref{eqn:WST}. B is arbitrarily limited to [-0.5; 0.5] and [0; 0.2 $\pi$] $\cup$ [0.8 $\pi$;$\pi$ ] for $\phi$. 

The capture range of this method is defined in \cite{Pinna:thesis} by the magnitude of the smallest detectable $OPD$ and the largest detectable $OPD$.
The smallest detectable $OPD$ is defined as: 
\begin{equation}
    OPD_{m}=\frac{\Delta \phi_{m}}{2 \pi} \frac{\lambda_{e} \lambda_{s}}{\lambda_{e}-\lambda_{s}}
    \label{eqn:SmallStep}
\end{equation}
With $\lambda_{e}$ the wavelength at the end or the biggest wavelength of the sweep range, $\lambda_{s}$ the wavelength at the start or the smallest wavelength of the range, $\Delta \phi_{m}$ the smallest phase change tolerated within the sweep range. 
The recommendation by \cite{Pinna:thesis} is to use $\Delta \phi_{m}=\pi$. 
This is equivalent to assuming that there should not be less than half a period between $\lambda_{s}$ and $\lambda_{e}$.
$OPD_{m}$ also describes the limit under which the random errors start to be an issue. 
$OPD_{m}$ is a consequence of the limited sweep range. 
When $OPD<OPD_{m}$, less than half of a period is present in the sweep range and a fit tries to find \eqnref{eqn:WST} in the noise of the signal. 
Consequently the $A$ parameter of \eqnref{eqn:WST} is of the order of the signal noise amplitude and $OPD$ takes a random value.

The largest $OPD$ is given by:
\begin{equation}
    OPD_{M}=\frac{\Delta \phi_{M}}{2 \pi} \frac{\lambda ( \lambda +\delta \lambda)} {\delta \lambda}
    \label{eqn:BigStep}
\end{equation}
With $\lambda=\lambda_{e}$, $\delta \lambda$ is the sampling of the sweep and $\Delta \phi$ the largest phase change tolerated between two samples.
Again, the recommendation by \cite{Pinna:thesis} is to use $\Delta \phi_{M} = \pi/2$.
This is equivalent to assuming in the worst case a quarter of a period between $\lambda_{s}$ and $\lambda_{s}+\delta \lambda$. 
This recommendation is a consequence of the Nyquist criterion which states that at least two points per periods are required. The Nyquist criterion says that there should not be more than half a period between two samples. We use a quarter of the period ($\Delta \phi_{M} = \pi/2$) meaning four points per period as a bare minimum.
The capture range of the method is $[-OPD_{M};-OPD_{m}]\cup[OPD_{m};OPD_{M}]$ or $OPD_{m}\leq|OPD|\leq OPD_{M}$ with $|OPD|$ the magnitude of the measured $OPD$.

Using \eqnref{eqn:BigStep} and \eqnref{eqn:SmallStep}, one can define the number of wavelengths required. \eqnref{eqn:SmallStep} helps define the total sweep range $\lambda_{e}-\lambda_{s}$. \eqnref{eqn:BigStep} helps define the sampling $\delta \lambda$. The minimum number of wavelengths required for the wavelength sweep, to give a reliable answer is given by $(\lambda_{e}-\lambda_{s}) /\delta \lambda$.

\subsection{Phase closure}\label{subsec:pc}

In \cite{PinnaSPIE:06,Bonaglia:10}, the wavelength sweep requires some error checking. 
The aim is to verify the consistency of all the OPDs measured and to find the sign errors and the random errors. 

This study's error check is based on the phase closure made for long baseline astronomical interferometry \cite{ObsAstr:12}. 
In interferometry, pupils are grouped by three and a super computer tries to nullify the phase delay for all groups of three pupils. Here instead of pupils, groups of three connected borders are made and the $OPD$s are summed. If the sum is zero, all three $OPD$s are consistent. 

On segmented mirrors, the groups of three borders are made by finding all borders that share a common end, e.g.: in \figref{fig:state}, the borders separating segments 1, 6 and 18 have a common end. For convenience, we call  groups of three borders a corner.
Each corner can be summed clockwise (this is important for the sign of the $OPD$). 
In an ideal case without any sources of noise, if their sum adds up to zero, then all three $OPD$s are correct. 
If they do not sum to zero, one or two of the $OPD$s are wrong. 
The wrong $OPD$ can be found by looking at the surrounding corners.

If another neighbouring corner's sum is also not zero, then the border between the two corners is the culprit.
There are two possibilities: a sign error or a random error. 
In the case of a sign error, the sign of the $OPD$ can be changed and the sum will become zero. 
In the case of a random error, changing the sign will not make the sum zero. 
The random error should be flagged and later, during the mirror segment's piston reconstruction, ignored.  

The algorithm stores which $OPD$s need to have an opposite sign to the one calculated and which $OPD$s need to be discarded, because of the random errors.
Some combinations of borders with sign errors or random errors render this strategy ineffective. 
The probability of occurrence of those combinations becomes too high if the number of detected wrong borders $OPD$s is >10-15\,\%.

In practice, because of noise, the sum of three borders never equals zero. 
The best found threshold  is: 
\begin{equation}
    T = 0.1 \sum_{i=1}^{3} |OPD_{i}|
\end{equation}
All corner sums less than $T$ are treated as if the sums are zero. This threshold minimizes the number of borders that are wrongly detected with an error, while it maximizes the number of borders correctly detected with an error. This is determined empirically with the help of the measurements made on the test bench. It is relatively high because the phase closure does not take into account the bias that accompanies a sign error (see section \ref{subsec:signError}).

\subsection{Zernike phase contrast sensor}\label{subsec:zeus}

The wavelength sweep relies on a phasing sensor to give it the information represented by the round markers on \figref{fig:WST2}. 
We use the Zernike phase contrast sensor to this end.

The Zernike Phase contrast sensor was first described in \cite{Dohlen:03,Surdej:10,Vigan:11}.
Its setup, in the context of the present work, is presented in \figref{fig:schem}. Schematically, it is a phase mask placed in the focal plane of the telescope, centred on the star image, and a re-imaging optic to project an image of the pupil plane on the detector. 
The signal of the phase mask is an intensity variation, perpendicular to the separation between each pair of two adjacent segments in the pupil plane.
The effect of the phase mask is to turn only the high spatial orders of the phase into an intensity signal. 
For a simple mathematical description, the reader is directed to \cite{hecht:book} section 13.2.4.
The phase mask is a parallel plate with a top hat circular depression at its centre. The depression has a radius $a$ and creates an optical path difference $D$. $a$ and $D$ should be tuned to the atmospheric conditions and the wavelength used.
We are working with a ground based telescope and, in a worse case scenario, without adaptive optics. Ideally, we would like to ignore the effect of the atmosphere.
$a$ has an influence on the cutoff frequency of the high pass filter and the width of the signal. 
$a$ should be bigger than half the atmospheric seeing, to remove as much as possible the lower spatial frequency wavefront distortions\cite{Surdej:10}. $a$ is defined in angular units to remain independent from the telescope plate scale.

\begin{figure}[h!]
    \centering
    \includegraphics[width = \textwidth]{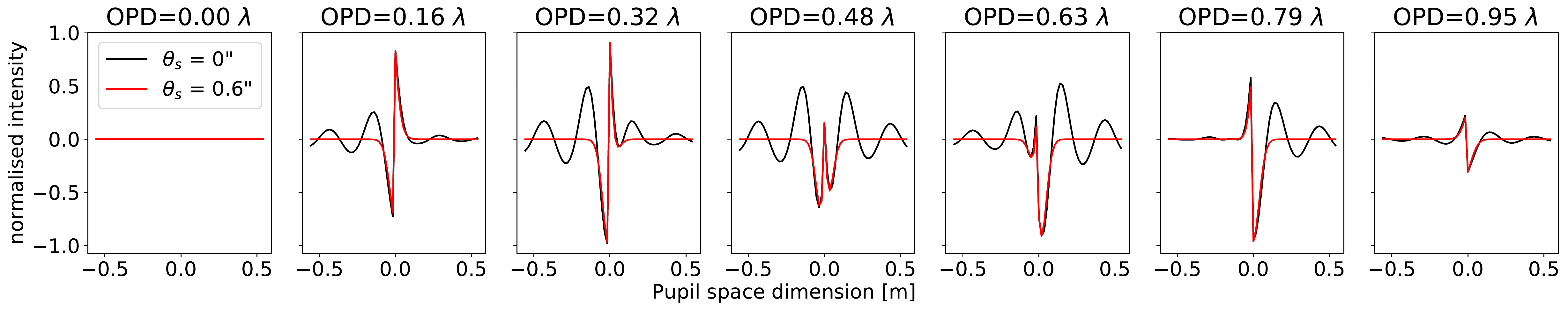}
    \caption{Simulation of the evolution of the Zernike phase contrast signal with increasing $OPD$ between two adjacent segments. The phase mask has a radius $a = 0.5\,"$ and a depth $D = \lambda/4$. The title of each plot gives the current relative height or $OPD$ between the two segments. The x-axis is projected on a VLT primary mirror. The black line represents the signal without atmosphere ($\theta_{s} = 0$) and the red line with a seeing of 0.6\," ($\theta_{s} = 0.6 "$). }
    \label{fig:evo}
\end{figure}

The phase contrast signal can be anti-symmetric across the border. This property gives the indication of which segment is higher than the other one, and therefore the direction of the correction.
The anti-symmetry arises only if $D$ is close to a quarter of the wavelength of observation. 
The signal becomes completely symmetric when $D$ is half a wavelength. 
Ideally, $D$ should be a quarter of the middle wavelength of the sweep range.
Because of the available hardware, this study uses a phase mask optimised for $\lambda =$700\,nm, $D=700/4$\,nm = 175\,nm. We define $D$ in metric units. It is converted to the wavelength dependent phase angle by the fitting algorithm.
In our case, we can also approximate $D$ as constant, as the refraction index of the materials used doesn't change significantly across the sweep range ($ \delta n_{SiO_{2}} = 0.006$ and $\delta n_{air} = 10^{-5}$). 
 \figref{fig:evo} shows the signal evolution with an increasing $OPD$ from left to right.

The extraction of the information needed for the wavelength sweep is done by a fit as described in \cite{Surdej:10} in section 3C. 
The signal can be described and fitted with a 6 parameters equation(\cite{Surdej:10} equation 21). 
\begin{equation}
    \begin{split}
        F(x) = & a_{3} + a_{4}(x-a_{5}) \\
        & + [1-f_{mask}(a_{6}(x-a_{5}))][a_{1} f_{step}(x-a_{5}) \sin{(\psi)} \\
        &- a_{2} f_{mask}(a_{6}(x-a_{5})) (1-\cos{(\psi}))]
    \end{split}
    \label{eqn:zeus}
\end{equation}
With:
\begin{equation*}
    \begin{split}
        f_{mask}(a_{6}(x-a_{5})) &= \frac{2}{\pi}\int_{0}^{a_{6}(x-a_{5})}\frac{\sin{(t)}}{t}dt\\
        f_{step}(x-a_{5}) &= \frac{e^{s(x-a_{5})}-1}{e^{s(x-a_{5})}+1}
    \end{split}
\end{equation*}
The $s$ parameter is adjusted once, depending on the system's resolution and gap. \cite{Surdej:10} omits to mention that in the actual fitting algorithm, the effect of the polishing errors and the gap are accounted for by replacing the $sign(x-a_{5})$ by $f_{step}$.
$a_{3}$ and $a_{4}$ are background terms that are mostly nullified by the image normalisation.
$a_{5}$ describes the signal centring on the border and is nullified by pupil registration. 
$a_{6}=\pi 2a/\lambda$ with $a$ the phase mask radius in radiant and $\lambda$ the wavelength in meters. 
This needs to remain a free parameter in the fit (within bounds) because it has slight dependency on the seeing conditions.
$\psi$ is the delay created by the phase mask in radian. $\psi=2\pi D/ \lambda $ where $D$ is the delay created by the phase mask in nm. 

To the first order, $a_{1}$ and $a_{2}$ are :
\begin{equation}
    \begin{split}
        a_{1} & = C_{1} \sin{(k OPD)}\\
        a_{2} & = C_{2} \cos{(k OPD)}
    \end{split}
    \label{eqn:a1}
\end{equation}
where $k$ is the wavenumber, $OPD$ is the current $OPD$ at the measured border.  $C_{1}$ and $C_{2}$ are two coefficients that describe mainly the influence of the atmosphere on the signal. 
The equation of $a_{1}$ can be directly linked to the \eqnref{eqn:WST} of the wavelength sweep and justifies the use of the phase contrast sensor.

For the wavelength sweep, since the wavelength changes between each measurement, the signal's shape changes too. There are two effects: like for any interferometer, the same $OPD$ does not represent the same fraction of the wavelength, hence the signal's magnitude changes accordingly. $a_{1}$ and $a_{2}$ describe this effect. The second effect is linked to the depth of the phase mask, which does not represent the same phase delay for each wavelength. This will change the evolution of the  of the signal's shape. $\psi$ in \eqnref{eqn:zeus} is the description of the changes in the signal shape. It controls whether the signal will be more anti-symmetric or symmetric with the term: $[a_{1} f_{step}(x-a_{5}) \sin{(\psi)}- a_{2} f_{mask}(a_{6}(x-a_{5})) (1-\cos{(\psi}))]$. If $\psi = \pi/2$ then the cosine part will be cancelled out. The signal will always look like the one in \figref{fig:evo} at $OPD = 0.16\lambda$ and will only change in magnitude. As $\psi$ changes toward $0$ or $\pi$, the cosine contribution will become more important, while the sine will decrease. When $\psi = 0$ or $\pi$ the signal will look like \figref{fig:evo} with $OPD = 0.48\lambda$. Only the amplitude will change.

A good fit to this equation is only possible if the following measures are taken. 
The phase mask's central depression needs to be centred on the star image, with a precision better than $a/8$. 
The pupil registration must be done with a precision better than the pixel size, for the outer-most ring of borders. 
Finally, the setup needs to be achromatic, e.g. to avoid defocus as a function of wavelength. 

\subsection{Overview of a measurement}\label{subsec:over}

\begin{figure}[h!]
    \centering
    \includegraphics[width=\textwidth]{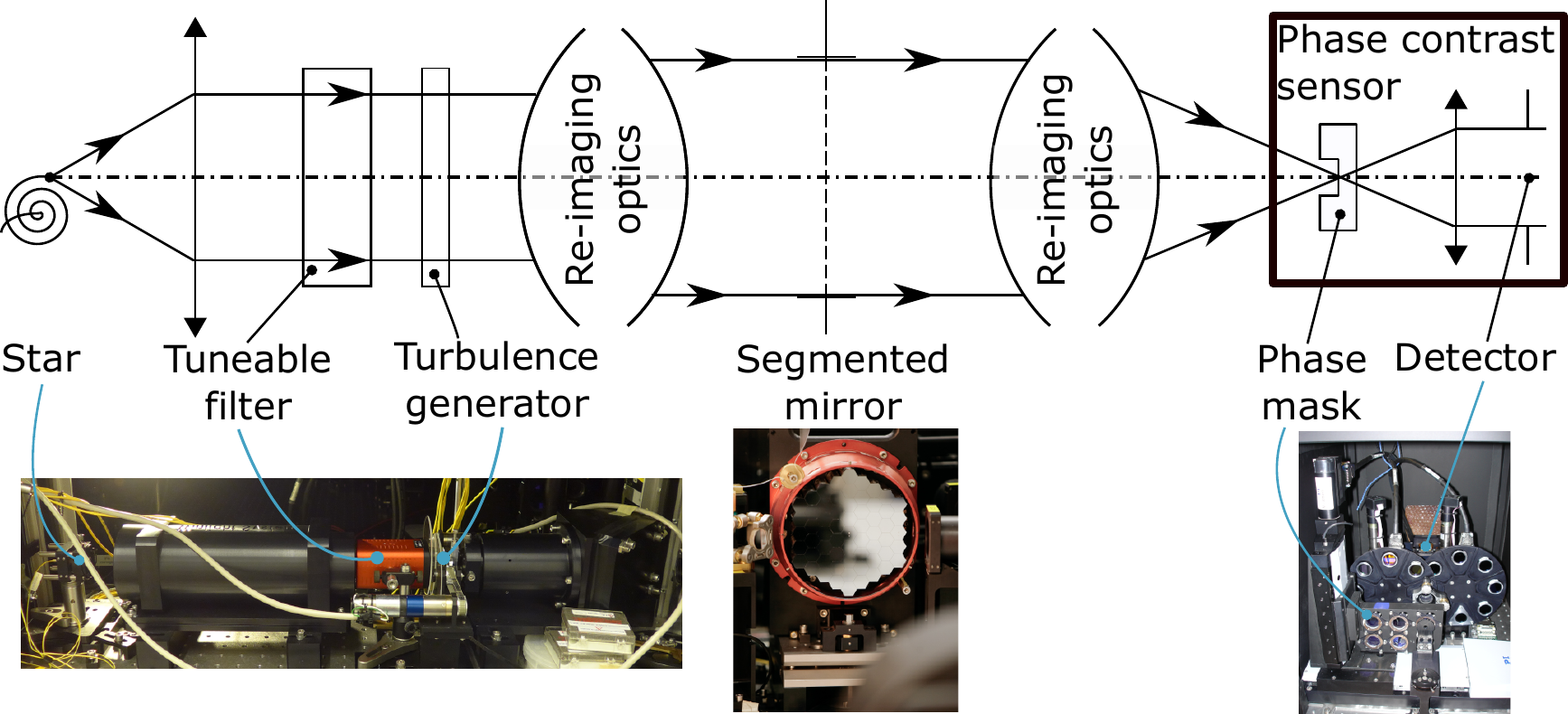}
    \caption{Schematic of the bench light path.}
    \label{fig:schem}
\end{figure}

\figref{fig:schem} shows schematically the test setup used to validate the wavelength sweep technique. 
The test bench starts with a fiber head source that emulates a star, which injects a white light into the bench. 
A lens collimates the light beam into the liquid crystal tuneable spectral filter\cite{lctfThor:ukn}. 
The turbulence generator MAPS\cite{Kolb:06} adds atmospheric turbulence to the beam. 
Re-imaging optics re-scale the beam, to cover the entire segmented mirror using. Another set of re-imaging optics focus the beam on the Zernike phase contrast sensor. Finally, a lens collimates it on the detector.

After setup, the internal metrology\cite{Wilhelm:08} drives the segmented mirror to a piston configuration, while images at different wavelengths are recorded.

\begin{figure}[h!]
    \centering
    \includegraphics[width=7cm]{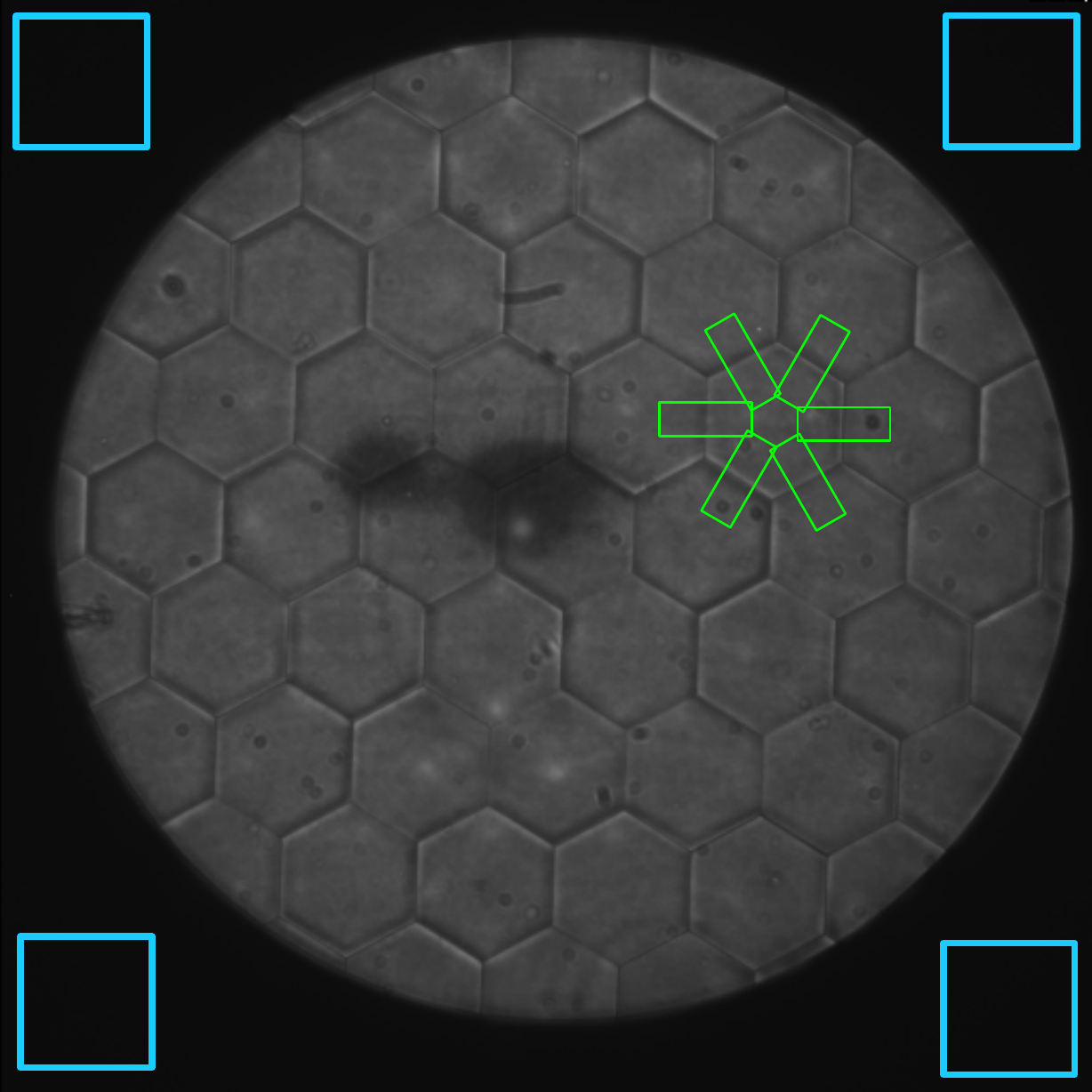}
    \caption{Typical image of the phase contrast signal in the pupil plane. The origin of the picture is in the lower left corner. The green boxes show an example of how the signal was extracted from the image. The blue square boxes indicate where the background is averaged from. The black peanut-shaped spot in the middle is a damaged area on the tuneable liquid crystal filter.}
    \label{fig:img}
\end{figure}

To use the phase contrast sensor, two images at each wavelength are required: one with the phase mask, the other one without\cite{Surdej:10}. 
The image without mask is only needed for normalisation of the image with the mask. 
The background for the normalisation is extracted from the corners of the image with the mask, as indicated in the blue boxes in Figure \ref{fig:img}. 
The normalised image with the mask is used to perform the pupil registration and to extract the phase contrast signal, as illustrated in \figref{fig:img} by the green boxes.

Each of the border's signals are first fit as described in section \ref{subsec:zeus}. 
The result of this fit at all the recorded wavelengths is then used to perform the wavelength sweep on each border, as described in section \ref{subsec:WST}. 
From the previous study \cite{PinnaSPIE:06,Bonaglia:10}, it is already identified that the wavelength sweep has two major errors: a sign error and a random error.
The sign error is a misdetection of the sign or direction of the OPD. 
This article explores the reasons for this to happen in section \ref{subsec:signError}. 
The random error happens when the OPD is too small. 
In response to these two errors, we perform a check on the result of the wavelength sweep and, when possible, correct  the sign errors as described in section \ref{subsec:pc}. 

Finally, the piston of each segment is computed, using the pseudo inverse of the synthetic interaction matrix linking the segment pistons with the border's OPD.

\section{Experimental results}\label{sec:result}

\subsection{Setup and data acquisition}\label{subsec:mdE}

\begin{figure}[h!]
\centering
\includegraphics[width=7cm]{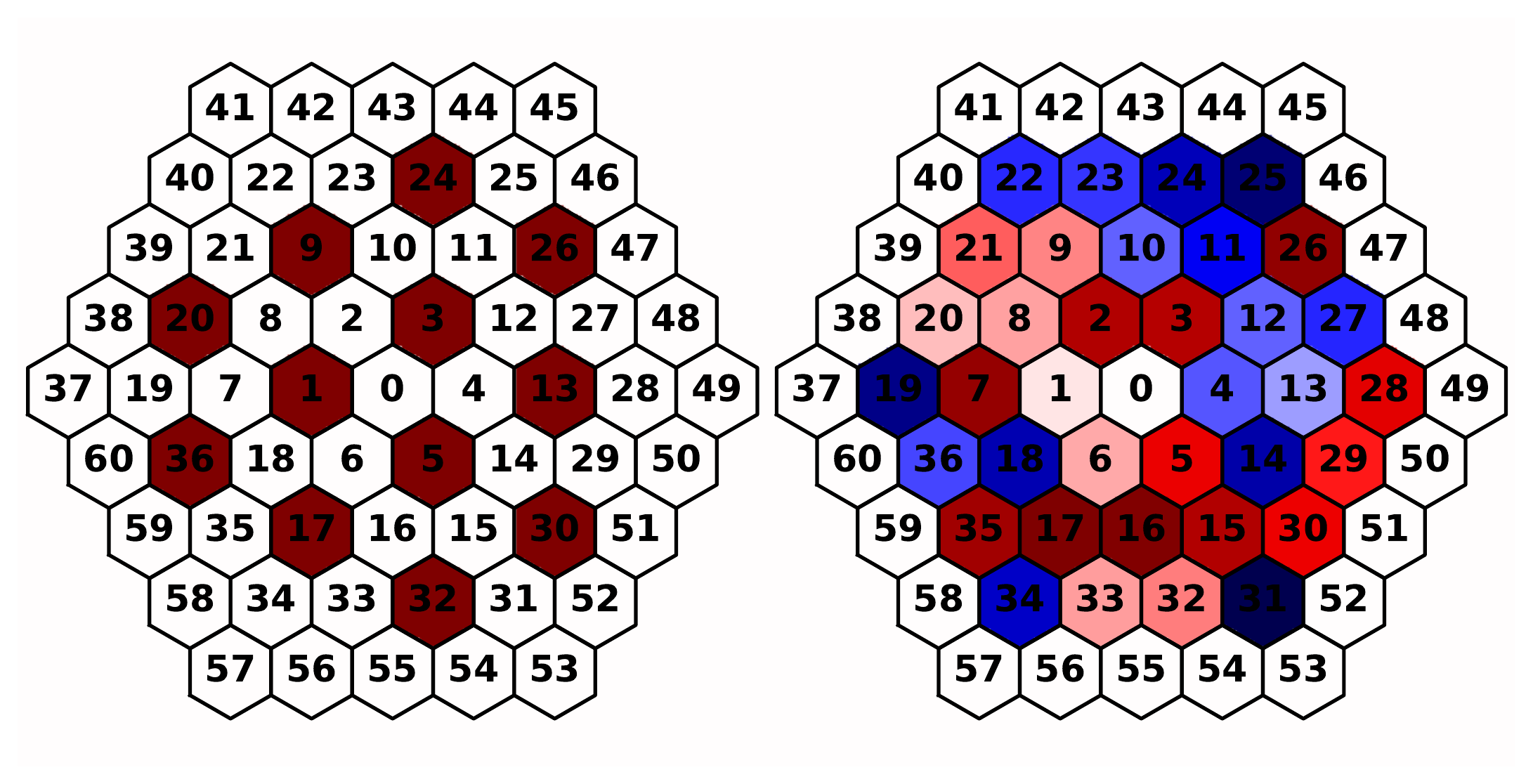}
\caption{Example of imposed piston configuration. Left: "family", has been repeated several times with all segments in red placed sequentially at -6\,000,-3\,000, 2\,000, 4\,000, 5\,400\,nm $OPD$. Right: "random" shows an example of one of the six random configurations applied. The distribution of the pistons is uniform and distributed between $\pm$6\,000\,nm, $\pm$4\,000\,nm and $\pm$2\,000\,nm. All segment positions are given relative to the centre segment 0. }
\label{fig:state}
\end{figure}

The goal of the experiment is to use the wavelength sweep method to measure the phasing state of the segmented mirror and determine the precision with which the piston could be measured in various scenarios. 
The segments are initialized into two types of configurations shown in \figref{fig:state}, "family" and "random". 

The family configuration puts all segments in red at the same piston. 
The random configuration applies a random piston to each segment. 
In different measurements, the red segments in the family state shown in \figref{fig:state} left took the pistons -6\,000, -3\,000, 2\,000, 4\,000, 5\,400\,nm. 
The state shown on the right is an example of the six random states used. 
The segments' piston is uniformly and randomly distributed between $\pm$6\,000\,nm for two configurations, $\pm$4\,000\,nm for two more and $\pm$2\,000\,nm for the final two.
Each of the presented configurations are measured separately. The result for all of them confounded are presented in the following sections.

In order to apply the family and random configurations, a reference measurement system called the internal metrology \cite{Wilhelm:08} is used. The internal metrology is a dual-wavelength Michelson interferometer, using a four points phase shift polarisation encoded scheme to retrieve the phase of each wavelength. One arm of the Michelson interferometer reflects off the segmented mirror, while the other one reflects off a reference mirror. The two wavelengths are alternatively shined in the interferometer and a synthetic wavelength algorithm is applied.

The choice of the wavelengths in the sweep is driven by \eqnref{eqn:SmallStep} and \eqnref{eqn:BigStep}.
The smallest $OPD$ detectable is $\approx$930\,nm because $\lambda_{s}$ = 650\,nm for the Liquid Crystal Filter. $\lambda_{e}$ = 1\,000\,nm due to the cut-off wavelength of the detector, setting the sweep range from 650 to 1\,000\,nm. $\Delta \phi _{m} = \pi/2$
For the largest $OPD$, the goal is to find the appropriate $\delta \lambda$ to cover the required capture range. 
The wavelength sweep needs to cover the capture range of our reference measurement method which is $\pm$6\,000\,nm. 
\eqnref{eqn:BigStep} can be rearranged as (Assuming $\Delta \phi _{M} = \pi/2$):
\begin{equation}
\delta \lambda = \frac{\lambda^{2}}{4 OPD_{M}-\lambda}
\end{equation}
Where $OPD_{M}$ is the required capture range. 
For $OPD_{M}$ = 12\,000\,nm, $\delta \lambda \approx$21.3\,nm. 
For convenience in measurements, $\delta \lambda$ is set to 20\,nm which makes an $OPD_{M}$ = 12\,750\,nm.

Finally, all data points at 830 and 850\,nm are removed, because of a notch filter in the beam between 815 and 865\,nm. 
The notch is present to cut out the wavelength used by the internal metrology, scattered by the segmented mirror.
In total 16 images at a different wavelength were taken, starting at 650\,nm by step of 20\,nm, up to 990\,nm. 

\subsection{Wavelength sweep result}
\begin{figure}[h!]
    \centering
    \includegraphics[width = 7cm]{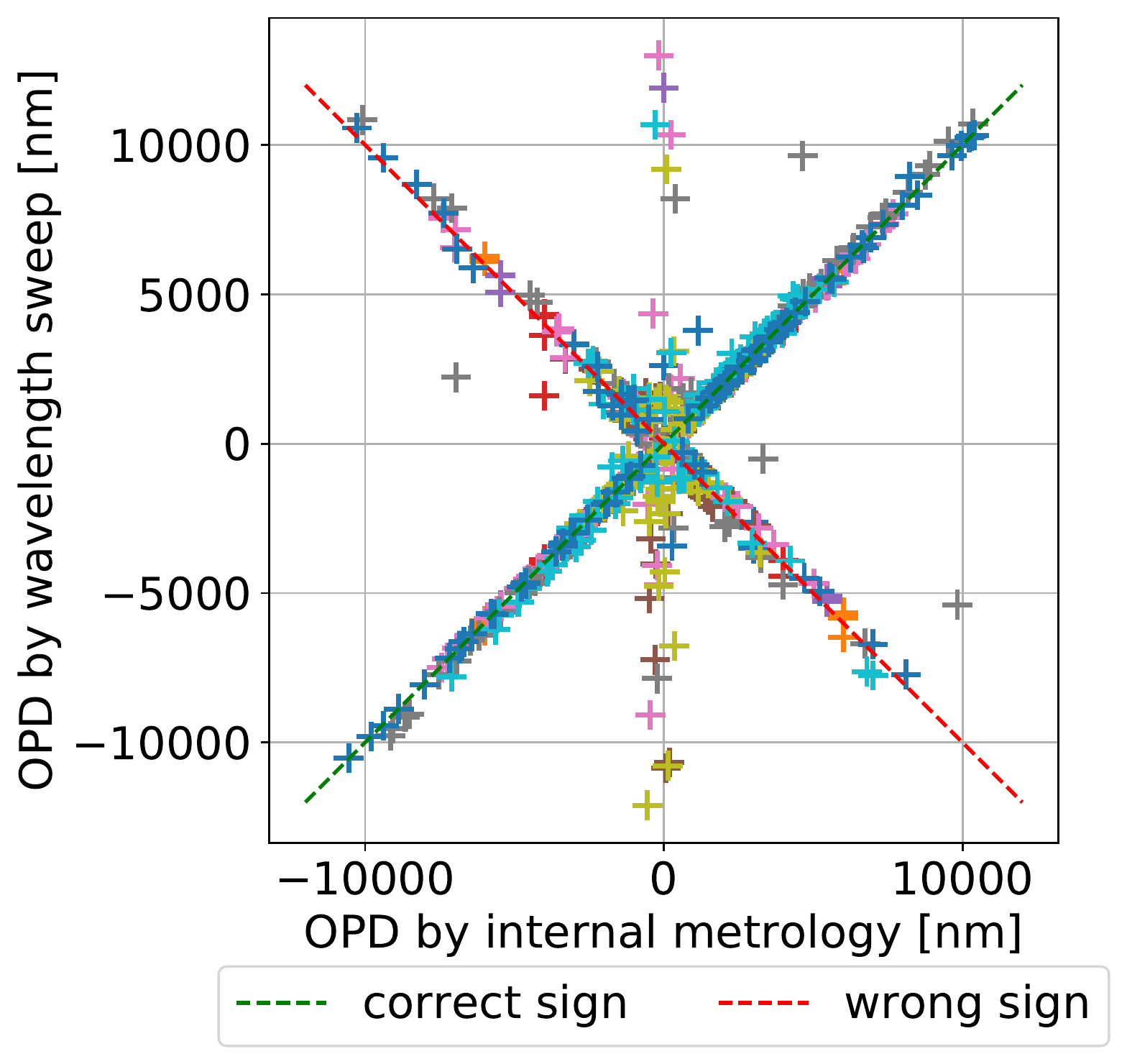}
    \caption{$OPD$ measured by the wavelength sweep versus the $OPD$ measured by the internal metrology. The different colours of the crosses denote each configuration (family -6\,000, -3\,000 ... and random $\pm$ 2\,000 ... etc.) that was applied to the segmented mirror. }
    \label{fig:cross}
\end{figure}

\figref{fig:cross} shows the measurement results of the wavelength scan for all the configurations (family and random) differentiated by crosses of different colours. 

In addition to the many correct measurements on the green dashed line, there are two interesting features in this result. First, all the data points that surround the red-dashed diagonal with a sign error, secondly closer to zero on the x-axis, all the data points with random errors. 
They correspond to the errors already reported by \cite{Pinna:thesis}.
The sign error is explored further in section \ref{subsec:signError}.

\begin{figure}[h!]
\centering
\includegraphics[width=7cm]{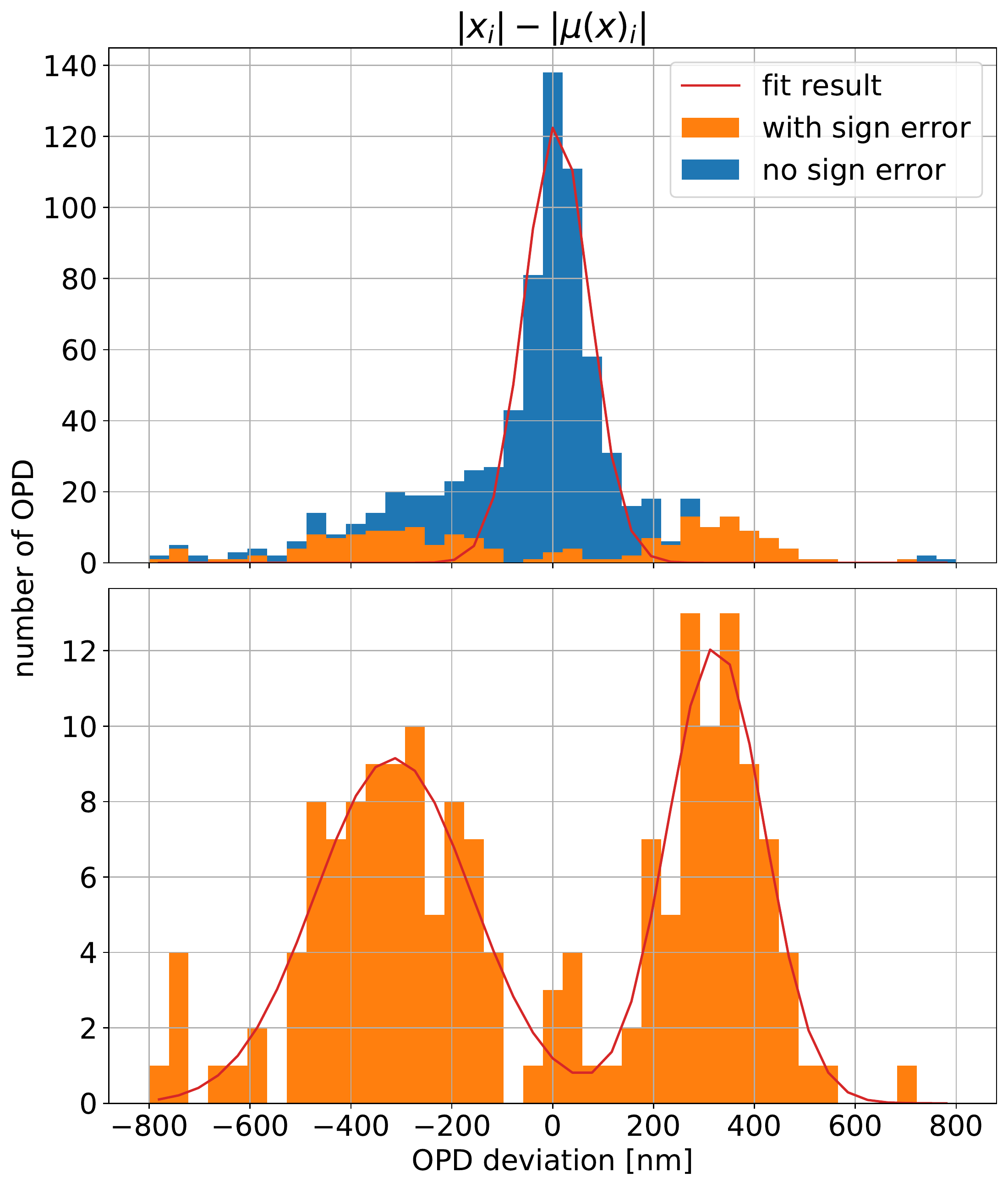}
\caption{The top histogram is the difference between the absolute value of the $OPD$ by the wavelength Sweep ($|x_{i}|$) and the absolute value of the $OPD$ by the internal metrology ($|\mu(x)_{i}|$). The blue bars are the data without a sign error and the orange ones these with a sign error. The bars are stacked. The bottom histogram shows only the data with a sign error compared to the Internal Metrology. The results of the gaussian fits to the histograms are shown in \tabref{tab:fitDat}.}
\label{fig:hisRes}
\end{figure}

The random errors are discriminated by a threshold on the parameter $A$ of the wavelength sweep and their $OPD$s are set to 0.
\figref{fig:hisRes} shows the precision of all the measured $OPD$s; the values are reported in \tabref{tab:fitDat}. 

\begin{table}[h!]
\centering
\caption{Result of fitting the histogram in \figref{fig:hisRes}.}
\begin{tabular}{|m{3cm}|c|c|}
\hline
\multirow{2}{*}{Figure} & \multicolumn{2}{c|}{fit result} \\ \cline{2-3} 
 &   $\mu$ (nm)   & $\sigma$ (nm) \\ \hline
\figref{fig:hisRes} top (only without sign error) &  8.54  &   64.50  \\ \hline
\multirow{2}{3cm}{\figref{fig:hisRes} bottom (only with sign error)} & -315.62  & 155.02 \\ \cline{2-3} 
                  & 323.86 & 95.74 \\ \hline
\end{tabular}
    \label{tab:fitDat}
\end{table}

The top histogram shows this for all $OPD$ measurements and the bottom one for all the measurements that have a sign error. The shape of the histogram with sign error is a consequence of the least square equation applied to this problem. A more detailed explanation is given in section \ref{subsec:signError}.

\figref{fig:hisRes} does not include all the $OPD$s of the "family" configuration (\figref{fig:state} left) that are phased. There are 30 phased borders measured five times. Only three borders were not detected as phased, including one border that was fitted twice as non-zero. 

The blue boxes in \figref{fig:hisRes} top, show the histogram of the deviation of the wavelength scan measurement with respect to the internal metrology. It is precise to 64\,nm RMS over all the measurements without sign error. This is to compare with the capture range of a monowavelength phasing sensor e.g.: \@ 650\,nm the capture range is $\pm$ 162\,nm. This already demonstrates the capacity of this method to reduce the phasing error to the capture range of a monowavelength phasing sensor.

\subsection{Interpretation of the sign error} \label{subsec:signError}

In \figref{fig:cross} the red dashed line represent the sign error. It affects on average 30\,\% of the $OPD$s in the present data set. 
The sign error does not systematically affect the same border in repeated measurements over the entire mirror. 
Borders that were identified as potentially problematic (e.g.: because of a dust too close to the border or damage to the optical surfaces) do not show a higher rate of wrong sign detection than others. 
We cannot explain the sign error with hardware problems.
The sign error is detected in up to 12\% of the $OPD$s in the family configuration of \figref{fig:state}. 
This number is better for the family configurations because 33\,\% of the borders do not have an $OPD$. In the 66\,\% of borders that have an OPD, 18\% have a sign error.
For the "random" configurations 23-50\,\% of the borders have a sign error.
A potential explanation for the higher number of sign errors in the random configuration is cross-talk between borders.

A property of the sign error is shown on the bottom histogram in \figref{fig:hisRes}.
It shows two groups of data points, bracketing the zero deviation point. 
These groups give a clue as to the origin of borders with incorrectly detected signs. 

This study uses a least square criterion to find the best solution:

\begin{equation}
    R = \sum_{i=1}^{N}(S(\lambda_{i})-M(\lambda_{i}))^{2}
    \label{eqn:lsm}
\end{equation}

\begin{figure}[h!]
    \centering
    \includegraphics[width = \textwidth]{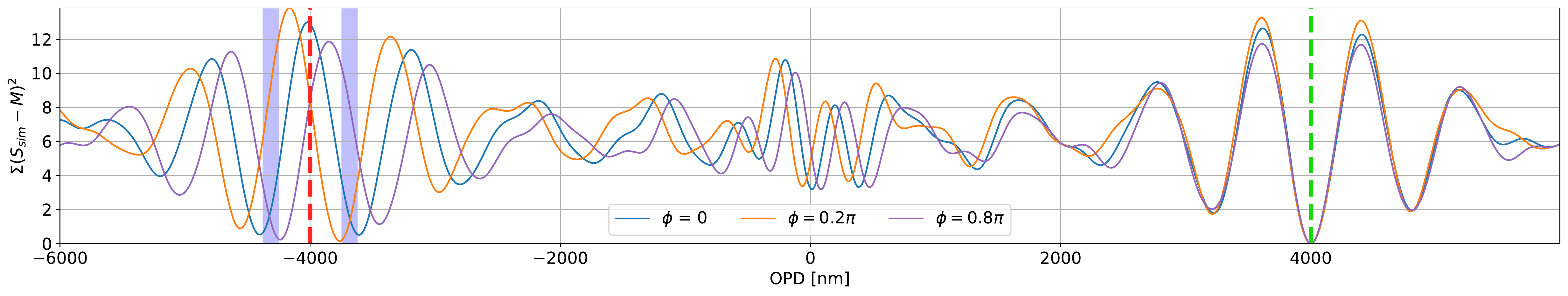}
    \caption{Evolution of the least square criterion with the model's $OPD$.The sum squared of the point by point difference is calculated. The blue areas represent all the possible second best solutions that the $OPD$ could take when $\phi \in [0.8\pi;\pi[$ for the left most blue area. For the right blue area, it is all the $OPD$ values that could be the best solution in case $\phi \in ]0;0.2\pi]$. Each blue, orange and purple curves are referring to a value of $\phi$ in \eqnref{eqn:WST} which is respectively $0$, $0.2\pi$ and $0.8\pi$ for the model and the simulated signal as indicated on the figure.}
    \label{fig:lsq}
\end{figure}

Where $S$ represents the signal and $M$ an ideal model. 
\figref{fig:lsq} shows the evolution of $R$ with the $OPD$ given on the x-axis for the model $M$ with $S$ having a fixed $OPD$ at 4\,000\,nm. 

For the readability of this demonstration the signal $S$ is ideal and has no noise. 
The green vertical line represents the true answer and the red one the true answer with a wrong sign. 
Looking only at the blue curve, the red line is surrounded by two minima, at 400\,nm $OPD$ from the true absolute value (with the opposite sign).  
Additionally, the fitting algorithm used for the wavelength sweep allows the parameter $\phi$ to vary in [0; 0.2 $\pi$] $\cup$ [0.8 $\pi$;$\pi$ ]. 
Taking the extremes, the orange and purple curves respectively represent what happens to $R$ when the model has the same $\phi$ of respectively 0.2$\pi$ and 0.8$\pi$. 
The $\phi$ of $S$ is also respectively 0.2$\pi$ and 0.8$\pi$. 
The left minimum comes as close as 240\,nm to the magnitude of the true value when $\phi$ = 0.8$\pi$ and the same for the right one when $\phi$ = 0.2$\pi$. 
This phase variation explains the width of the two peaks in the bottom histogram of \figref{fig:hisRes}.

A physical reason for $\phi$ to drift can be found in \cite{Surdej:thesis} section 7.1.2: They show empirically that the alignment of the phase mask on the star's image has an influence on the phase contrast signal shape. In their model they translate the effect of the alignment in the equation of $a_{1}$ (to compare with \eqnref{eqn:WST}):
\begin{equation}
    a_{1} = A \sin{\left( \frac{2 \pi}{\lambda}OPD+\delta \right)} - A \sin{(\delta)}
    \label{eqn:a1better}
\end{equation}
Where $\delta\approx2.8\frac{w}{2a}$ with $a$ the radius of the mask and $w$ the distance of the star's image to the centre of the phase mask depression. $A$ and $OPD$ are the same as in \eqnref{eqn:WST}.
For $\delta$ to be of the order of $0.2 \pi$, the star's image needs to be $0.14a$ off the mask centre. 
The formula suggests that this misalignment should produce over all signals an offset of $A \sin{\delta}$. 
In the data, a corresponding offset signature is only found in 15\,\% of all $OPD$ measurements. 
According to \cite{Surdej:thesis}, a mask misalignment should create a consistent offset on all $OPD$s. 
Because this offset does not appear consistently across all borders, the misalignment cannot be responsible alone for the  sign errors.

Further simulations are required to explain the drift in $\phi$, this will be addressed in section \ref{sec:sim}. 
However, \figref{fig:lsq} allows us to define the noise threshold at which this drift of $\phi$ results in sign errors in the fit. 
There is a relation between the standard deviation and the minimisation criterion. 
The standard deviation is written as:

\begin{equation}
    \sigma = \sqrt{\frac{\sum\limits_{i=1}^{N} (x_{i}-\bar{x}_{i})^{2}}{N}}
    \label{eqn:std}
\end{equation}

With $x$ all the measurements and $\bar{x}$ their expected values from the model. 
\eqnref{eqn:lsm} and \eqnref{eqn:std} can be combined by defining $x=S(\lambda_{i})$, $\bar{x}=M(\lambda_{i})$ and $N$ the number of wavelengths used. 
One can write:
\begin{equation}
    \sigma = \sqrt{\frac{R}{N}}
    \label{eqn:noise}
\end{equation}
where $\sigma$ is a description of the average noise per point. 
To explain the number of sign errors, we investigate the impact of the noise on whether the fit finds the correct minimum or secondary minimum, using \figref{fig:lsq}. 
We find that when $\phi$ is 0 (blue curve) , a noise of 0.17\,x\,intensity will result in finding the secondary minimum. For $\phi = 0.2\pi$ (orange curve), a noise at roughly half that level will result in finding the secondary minimum. The same occurs for $\phi = 0.8\pi$ (the purple curve).
The average standard deviation on the bench is 0.13. This is in between the limits of 0.17 and 0.09 and it makes a drift in $\phi$ a plausible explanation for the number of sign errors. Again the only physical reason for $\phi$ to drift would be a misalignment between the star image and the phase mask. We will see in section \ref{ssec:simRes} that alternatively the fitting noise can also create drifts in $\phi$.
With lower noise, the fit would be robust against drifts in $\phi$.

\begin{table}[h]
    \centering
    \caption{Table of wavelength sweep performance for each piston configuration. The applied segment configuration is in the first column. The second column shows how many out of 90 borders have been detected with a wrong sign (the percentage of borders concerned). The third column indicates how many of the incorrect signs could be corrected using the phase closure algorithm. The fourth column gives the same information but in terms of percentage of borders that were successfully corrected.}
    \begin{tabular}{|c|m{1.5cm}|m{1.5cm}|m{1.5cm}|}
    \hline
    configuration & number of sign error  & overall corrected sign error & percentage of sign errors corrected\\
    \hline
    family -6000nm  & 11 (12\%) & 11& 100\%\\
    family -3000nm & 11 (12\%) & 11& 100\% \\
    family 2000nm & 0 (0\%) & 0& - \\
    family 4000nm & 9 (10\%) & 9& 100\% \\
    family 5468nm & 11 (12\%) & 5 & 45\%\\
    random 1 & 45 (50\%) & 3 & 6\% \\
    random 2 & 29 (32\%) & 8 & 28\%\\
    random 3 & 21 (23\%) & 8 & 38\%\\
    random 4 & 25 (27\%) & 8 & 32\%\\
    random 5 & 24 (26\%) & 13 &54\% \\
    random 6 & 32 (35\%) & 17 & 53\%\\
    \hline
    \end{tabular}{}
    \label{tab:signerror}
\end{table}

\subsection{Phase closure}
After measuring the $OPD$s, the phase closure algorithm attempts to correct any sign errors and detects remaining random errors.
\tabref{tab:signerror} presents per configuration the number of sign errors detected in the second column and how the phase closure algorithm performed in the third and fourth column.
\tabref{tab:signerror} confirms the threshold of 10 to 15\,\% of sign errors, established in Section \ref{subsec:pc}, for good performance of the phase closure algorithm.
Only the family configurations with 12\,\% and less sign errors are successfully corrected. 
The phase closure algorithm does not manage to correct enough $OPD$ in the random data sets. 
The random data sets output data are unusable for the rest of this study. 
  
The correction of configuration family 5\,468\,nm does not work in our test, because a random error could be compensated by two other $OPD$ having their signs changed. 
One of these two borders is at the edge of the segmented mirror, and hence does not cause more damage. The other border is not so conveniently placed and its sign change is compensated through other sign changes, resulting in a cascading error. 
In the data set family -6\,000\,nm and 4\,000\,nm one border in each was measured with a random error. 
Both borders are marked to discard by the phase closure algorithm. 
This is confirmed by looking at the fitted parameter $A$ of each border which was slightly above 0.1. 
As a result, the phase closure can also act as a second detection for random errors that passed the previous filter.

The family data sets show that the phase closure algorithm can help detect and correct for incorrectly measured $OPD$s. 
The random data sets also show that it is easily overwhelmed in the presence of many sign errors and random errors. 
One way of improving the phase closure algorithm would be to record during the wavelength sweep fitting the three best solutions for the $OPD$. 
The phase closure algorithm would try all possible solutions for all corners that have been calculated as not null. 

\subsection{Final mirror reconstruction}

Finally, the corrected $OPD$ measurements serve to reconstruct the piston of each segment, through the pseudo inverse of a synthetic interaction matrix. 
The result can be compared to the measurement of the internal metrology. 

\begin{figure}[h!]
    \centering
    \includegraphics[width = 7cm]{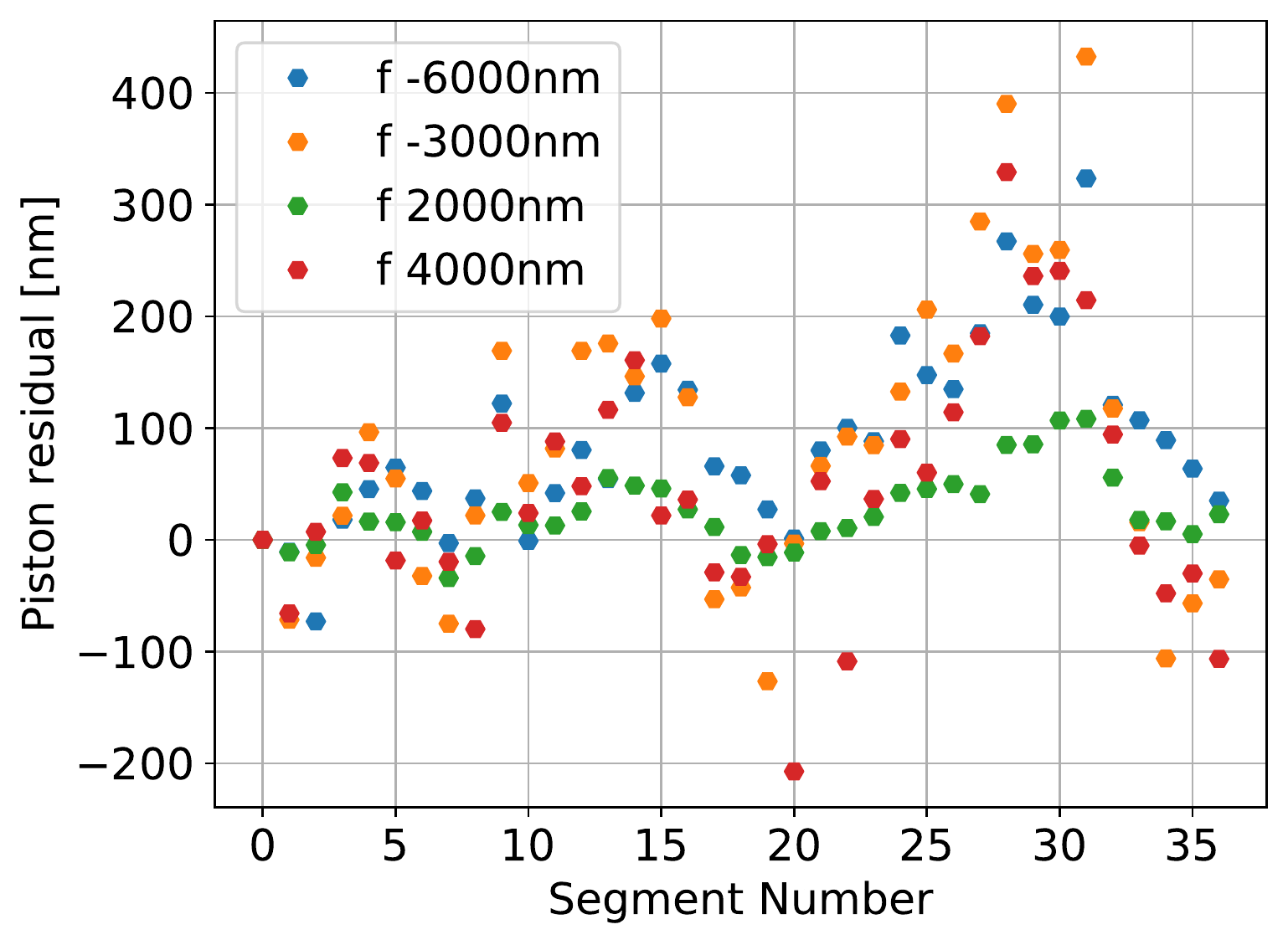}
    \caption{Residuals of the wavelength sweep measurement with respect to the internal metrology measurement.}
    \label{fig:corrStates}
\end{figure}

For each segment and for the data sets where the phase closure algorithm is successful, \figref{fig:corrStates} shows the residual after subtracting the wavelength sweep reconstruction from the internal metrology measurement.  
The precision of the reconstruction is 112\,nm RMS. 
The measurements from the configuration family 2\,000\,nm are interesting because there is no sign error to correct. 
This reconstruction is precise to 42\,nm RMS, nearly 3 times better than the average. 
The three other configurations are worse, because the sign correction did not try to correct for the $\approx$300\,nm bias that comes with the sign error. 
The idea of allowing the phase closure algorithm to choose from the three best $OPD$ measurements, when attempting to correct the sign errors would also correct this bias, potentially further improving performance of the mirror reconstruction in the presence of sign errors. 

The bias accompanying the sign error is even more relevant, if considering the aim of the wavelength sweep. 
We want to give the mirror a phasing state that is in the 100\,nm range, in order to do mono-wavelength phasing. 
If the PV of the residual piston is in the 400\,nm range, it means that we will need to use 1\,600\,nm as a wavelength for the mono-wavelength phasing. 
This is possible and sufficient for observations in the infra-red but not for observation in the visible. 
If the piston PV is of the order of  100\,nm, this means that working at 400\,nm for the monowavelength phasing can be achieved. 
This becomes excessive for most visible light observations, but acceptable for  UV observations.
 
\section{Simulations}\label{sec:sim}
\subsection{Model description}
The objective of this simulation is to find the worst offenders in the wavelength sweep, using the phase contrast. 
The criteria are: the number of $OPD$s with a sign error, the precision and accuracy of the fit, and the errors on small $OPD$ amplitudes.

This simulation is an image simulator and a simplified pupil registration algorithm. 
The output of the pupil registration algorithm is then fed into the fitting algorithm. 
The image simulator takes a 4096x4096 pixel input pupil wavefront. The input wavefront includes by default a segmented pupil with the possibility to add any piston to any segments. 
A series of operations is applied to this image:
(i) The input pupil is Fourier transformed into the focal plane, (ii) the focal plane is multiplied by the phase mask function, (iii) the resultant wavefront is then Fourier transformed to the output pupil plane, (iv) the sum of the square modulus of the output wavefront is calculated to create the image. (v) the image is recalculated for each wavelength of the spectrum's bandwidth. 
The simulation process is illustrated in the Algorithm~\ref{alg:imgSim}.

\begin{algorithm}
\caption{Pseudo code of the image simulator}
\label{alg:imgSim}
\begin{algorithmic} 
\STATE Set up all the simulation's parameters
\FOR{each wavelength of the sweep}
\STATE Create an empty image 
\FOR{each second of exposition}
\STATE Create a new atmospheric screen
\FOR{each wavelength of the LCTF bandwidth}
\STATE Create empty wavefront
\STATE Add to the wavefront the segment positions with positionning errors
\STATE Add to the wavefront the segment shapes
\STATE Add to the wavefront the Atmosphere
\STATE Add to the wavefront the defocus
\STATE Fourier transform of the wavefront
\STATE Add to the wavefront the phase mask with misalignment
\STATE Reverse Fourier transform of the wavefront
\STATE Add |wavefront|$^{2}$ to the image
\ENDFOR\COMMENT{each wavelength of the LCTF bandwidth}
\ENDFOR\COMMENT{each second of exposition}
\STATE image is binned
\STATE Add photon shot noise to image
\STATE Save the image
\ENDFOR\COMMENT{each wavelength of the sweep}
\STATE Output to the pupil registration algorithm
\end{algorithmic} 
\end{algorithm}

The Liquid Crystal Tunable Filter we are using to select the wavelength of observation has a bandwidth of roughly 10\,nm to 20 \,nm. The effect of the bandwidth can be simulated by adding the intensities at different wavelengths inside the bandwidth. In the current simulation, the final image is the sum of seven wavelengths: 
\begin{equation}
    I = \sum_{n=-3}^{3} |A_{n} \Psi(\lambda + n BW/2)|^{2}
\end{equation}
$A_{n}$ is the amplitude of the complex wavefront that is chosen according to a Gaussian distribution. 
$\Psi$ is the complex exponential part of the wavefront and $BW$ is the full width at  half maximum of the source bandwidth. 
All seven images are summed  to have the final image. 
Finally, this image is binned to match the 512x512 pixel camera resolution on the experimental test bench. 
By default, a photon shot noise is added to the final image, though this can be turned off for testing.

There are eight known sources of  noise : 1/ the readout noise, 2/ the variation of the spectral bandwidth of the light, 3/ the atmosphere emulated by MAPS, 4/ the segment positioning error of the internal metrology, 5/ a chromatic defocus, 6/ a misalignment of the mask, relative to the average star's image position, 7/ the segments misfiguration error and 8 / the pupil registration. The way each of these noises is added and their order is illustrated in the Algorithm \ref{alg:imgSim}.

The Liquid Crystal Tunable Filter provided by Thorlabs \cite{lctfThor:ukn}, has a broadening of the bandwidth with an increase in central wavelength. 
From the data provided by Thorlabs we can find a polynomial relation that describes this effect with respect to the central wavelength. The equation is:
\begin{equation}
    BW = 10^{-5}\lambda^{2} + 0.0019\lambda + 6.8659
\end{equation}
with $\lambda$ as the central wavelength in nanometres.

In the input complex wavefront, the noise effects that are added are the atmosphere, the positioning error, the segments' misfiguration and the chromatic defocus. 
On the bench, the atmosphere is simulated by MAPS\cite{Kolb:06}. 
The simulation reproduces its properties.
Because the phase contrast works better with long exposures, this is simulated by recalculating several times the output complex wavefront with as many different atmospheric screens as seconds of exposure time. 
30 seconds of exposure are used on the bench, hence 30 atmospheric screens are used for the simulation. 
The phase contrast being only sensitive to the high-frequency wavefront deformation, in 1 second they have changed enough to be approximately uncorrelated.

The internal metrology positioning error is translated into a Gaussian distribution of errors added to the position of the individual segments in the input complex wavefront. 
This error is an accumulation of the internal metrology measurement error\cite{Wilhelm:08} on the positioning system of the test bench \cite{APESPIE:08}. A previous study \cite{APESPIE:08} mentions that the RMS on the $OPD$ error is 0.75\,nm over 5 hours. 
Measurements done for this study show an RMS closer to 5\,nm RMS, probably due to ageing of the hardware. 
For each input wavefront, a new draw of piston errors with an RMS of 5\,nm is added.

The individual segment misfiguration is an information recorded by the internal metrology.
It shows a maximum of 30\,nm PV misfigure error. 
This information is added to each input wavefront on each segment, as the phase contrast is sensitive to the segment misfigure. 

The chromatic defocus seems to come from the last lens collimating the beam on the detector, according to the Zemax design of the test bench. 
According to the Zemax design, the defocusing at the detector as a function of the wavelength is: 
\begin{equation}
    D_{\lambda} = 0.5793\lambda^{2}-1.2791\lambda+0.5458
    \label{eq:defoc}
\end{equation}
With $\lambda$ in micrometers. 
Since the focus is set with a wavelength of 650\,nm, The defocus created on the camera will be $D_{\lambda}-D_{\lambda = 650}$

In the focal plane the misalignment of the phase mask relative to the position of the star image is added. 
An important hypothesis here is that the tip tilt of each segment is zero. 
On the test bench, this hypothesis is valid due to the internal metrology. 
Then we can describe the difference between the phase mask centre and the centre of the star image. 
On the bench, the alignment was done better than a quarter of the mask radius at 650\,nm. 
No verification has been made at other wavelengths and the initial alignment was done using the turbulence generator with a long exposure. 
For comparison the effect of misalignment has been evaluated with errors from 0 to 0.5 times the phase mask size.

The read out noise is added onto the rescaled output image, because it occurs at readout of the CCD pixels. 
Prototypes of the VLT Technical CCDs are used as cameras. 
These CCDs have 80\,$e^{-}$ RMS of readout noise. 
In addition, the readout noise presents structures that can be described by reconstructing the way the detectors are read. 
CCDs are read pixel by pixel, but the pixels need to be shifted to single readout output, which is located in one of the corners of the CCDs. 
The pixel matrix can be flattened by putting each row in a single line and then considered as a time series of intensities. 
According to documentation, a pixel is read each microsecond and the time needed to shift the next row in place is negligible. 
Dead pixels are replaced by an average of the full image without dead pixels. 
This time series is Fourier transformed and shows the frequency noise structures. 
The frequencies and amplitudes of the five highest peaks in amplitude are recorded and used to simulate this same configuration on the simulated images. 
Because these five frequencies can shift, this is done for 9 different darks to have a small statistical sample of the possible frequency shifts. 
\tabref{tab:noises} lists the recorded spectral results.

The last possible source of error is the pupil registration. 
The aim of this procedure is to retrieve the orientation of the hexagonal array, the size of the segments and the global centre of the hexagonal array. 
For the simulation, the pupil registration is done by giving the required pupil registration parameter. 
There are four parameters to adjust: the pupil orientation, the segments' size and the coordinates of the centre of the pupil. 
For the data these four parameters are found using the procedure described in \cite{Surdej:07}.
According to this article, the precision on the orientation is 12', 0.07 pixels for the segments' size and less than 0.07 pixels for the x and y position of the pupil centre. 
In order to test the robustness of the fitting to these errors, the orientation was tested from 1' up to 32', segment sizes from 0 to 0.5 pixels and centering from 0 to 2 pixels.


\begin{table}[h!]
    \centering
    \caption{list of sources of noise and their tested quantities}
    \begin{tabular}{|r|l|}
        \hline
        source of noise & characteristic \\
        \hline
        \multirow{6}{*}{Readout noise} & bias offset:1960 ADU ,\\
                      & $f_{1}$=20\,kHz, Amp = 1.9\,ADU \\
                      & $f_{2}$=360\,kHz, Amp = 1.8\,ADU\\
                      & $f_{3}$=340\,kHz, Amp = 1.4\,ADU\\
                      & $f_{4}$=320\,kHz, Amp = 1.0\,ADU\\
                      & $f_{5}$=40\,kHz, Amp = 0.8\,ADU\\
                      \hline
        positionning error & 10\,nm.$s^{-1}$ RMS\\
        \hline
        segments misfigure & max 30\,nm PV\\
        \hline
        \multirow{2}{*}{atmosphere @ 650\,nm} & layer 1 $r_{0}= 0.54$\,m \\
         & layer 2 $r_{0}= 0.33$\,m\\
        \hline
        \multirow{2}{*}{light bandwidth} & BW = $10^{-5}\lambda^{2} + 0.0019\lambda$ \\
          &  + 6.8659\\
        \hline
        phase mask  & 0.1, 0.2, 0.3, 0.4, 0.5\,mask diameter\\
        misalignement & \\
        \hline
        \multirow{3}{*}{defocus} & $D_{\lambda}-D_{\lambda = 650}$ with \\
         & $D_{\lambda}  = 0.5793\lambda^{2}$\\
         &                $-1.2791\lambda+0.5458 $\\
         \hline
        \multirow{5}{*}{pupil registration} & centering 0.1 to 1 pixel\\
        & by step of 0.1\\
        & segment size: 0.1 to 0.5 pixels \\
        & by step of 0.1\\
        & rotation : 1'to 32' \\
        & by step of 2'\\
        \hline
    \end{tabular}
    \label{tab:noises}
\end{table}
\subsection{Simulation results}\label{ssec:simRes}

In the simulation, with no added sources of noise, there are no sign errors. 
On the other hand under $\approx$1\,000\,nm, the simulation shows already random errors. This result includes light with a constant bandwidth of 10\,nm. The random errors observed would be reduced if the bandwidth was reduced.
This remains unchanged when the readout noise, the internal metrology positioning error, the broadening of the spectral bandwidth and the segments' misfigure are summed. 

However, the addition of the atmosphere generated by MAPS to all the above-mentioned effects creates on average three borders with a sign error. 
It also generates on average four borders with random errors. 
An image defocus added to the previous effect creates three borders with a sign error.
The phase mask misalignment can create up to 15 borders with sign errors on average when the misalignment is 0.5 of the phase mask radius. 
A misalignment by 0.1 mask-radius can create up to two wrong signs. 

Finally, including the pupil registration with centring errors and segment size errors, no additional errors are created. 
The misdetection of the orientation starts creating sign errors at 25' in the rotation. 

\begin{figure}[h!]
    \centering
    \includegraphics[width = 7cm]{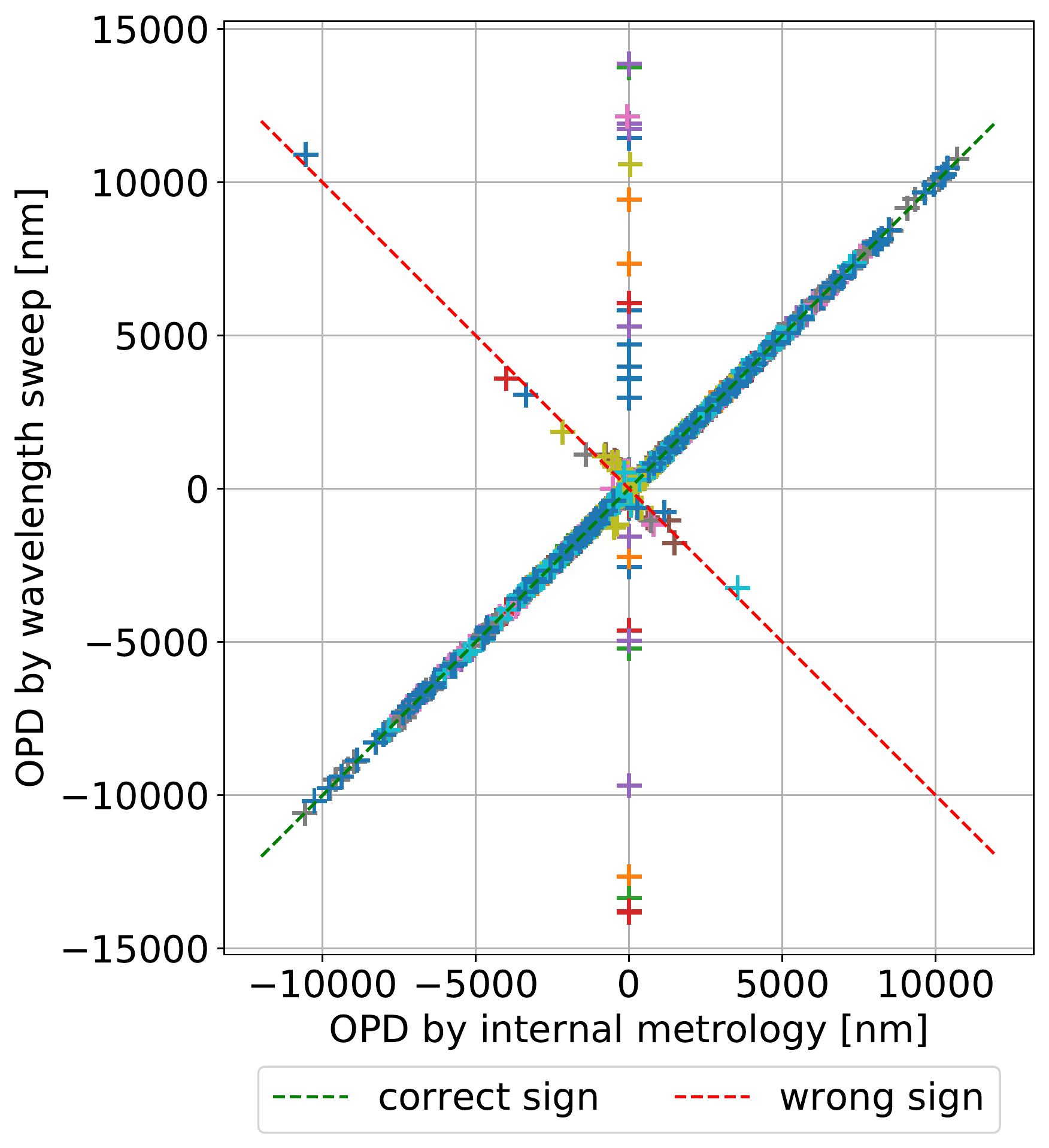}
    \caption{$OPD$ by the wavelength sweep versus imposed $OPD$. The different colors of the crosses denote each configuration that is applied to the  segmented mirror. These crosses show the result of the simulation with the noise contributions expected on the measurement test bench.}
    \label{fig:simcross}
\end{figure}

\begin{figure}[h!]
    \centering
    \includegraphics[width = 7cm]{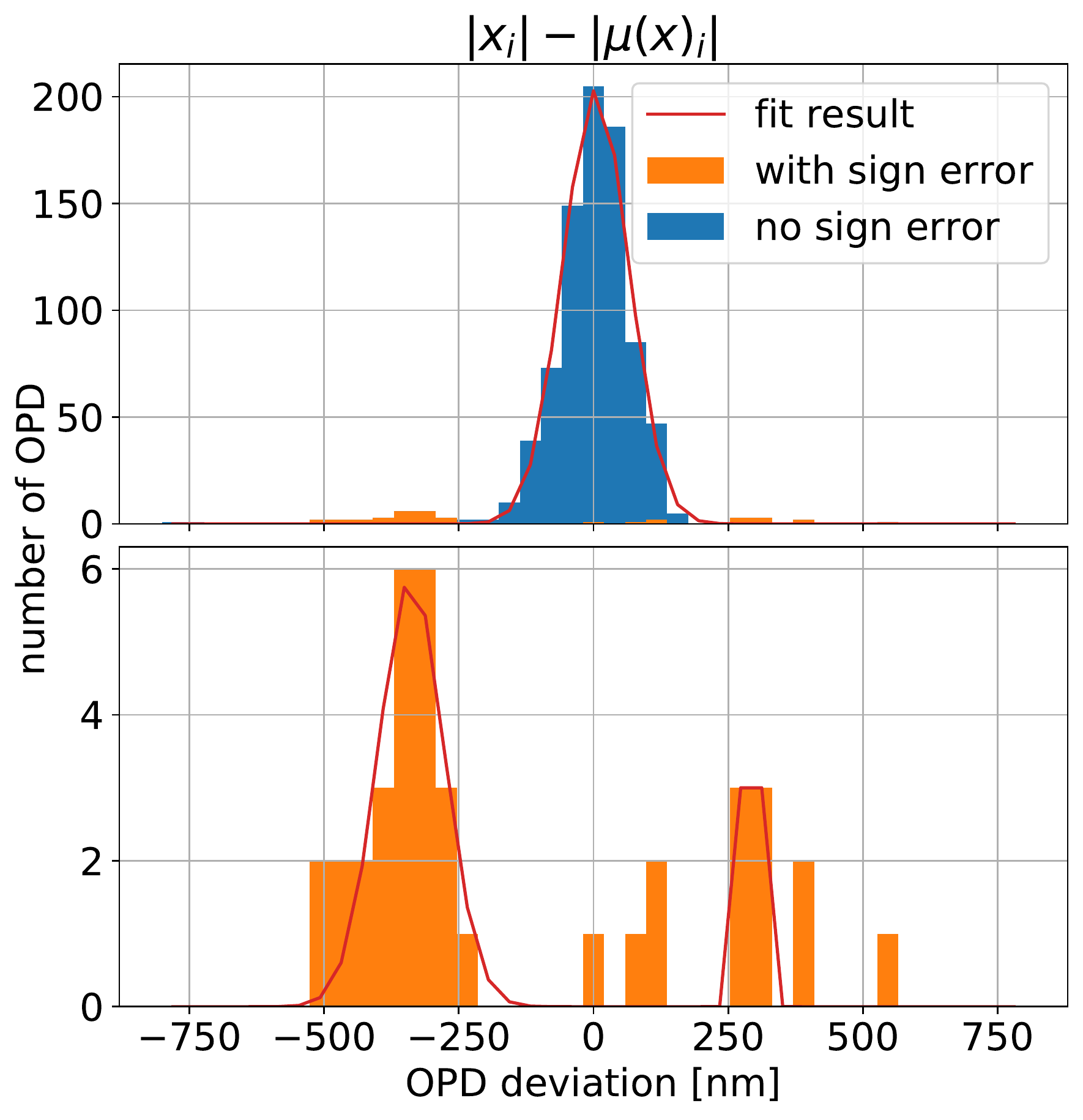}
    \caption{Distribution of the difference between the absolute value of the imposed $OPD$ and the absolute value of reconstructed $OPD$. The top diagram shows the histogram  for all data together and the bottom one only for the data with known sign error. The results of both fits are listed in \tabref{tab:fitSim}. In the top diagram with and without sign error are stacked.}
    \label{fig:simhist}
\end{figure}

\begin{figure}[h!]
    \centering
    \includegraphics[width = 7cm]{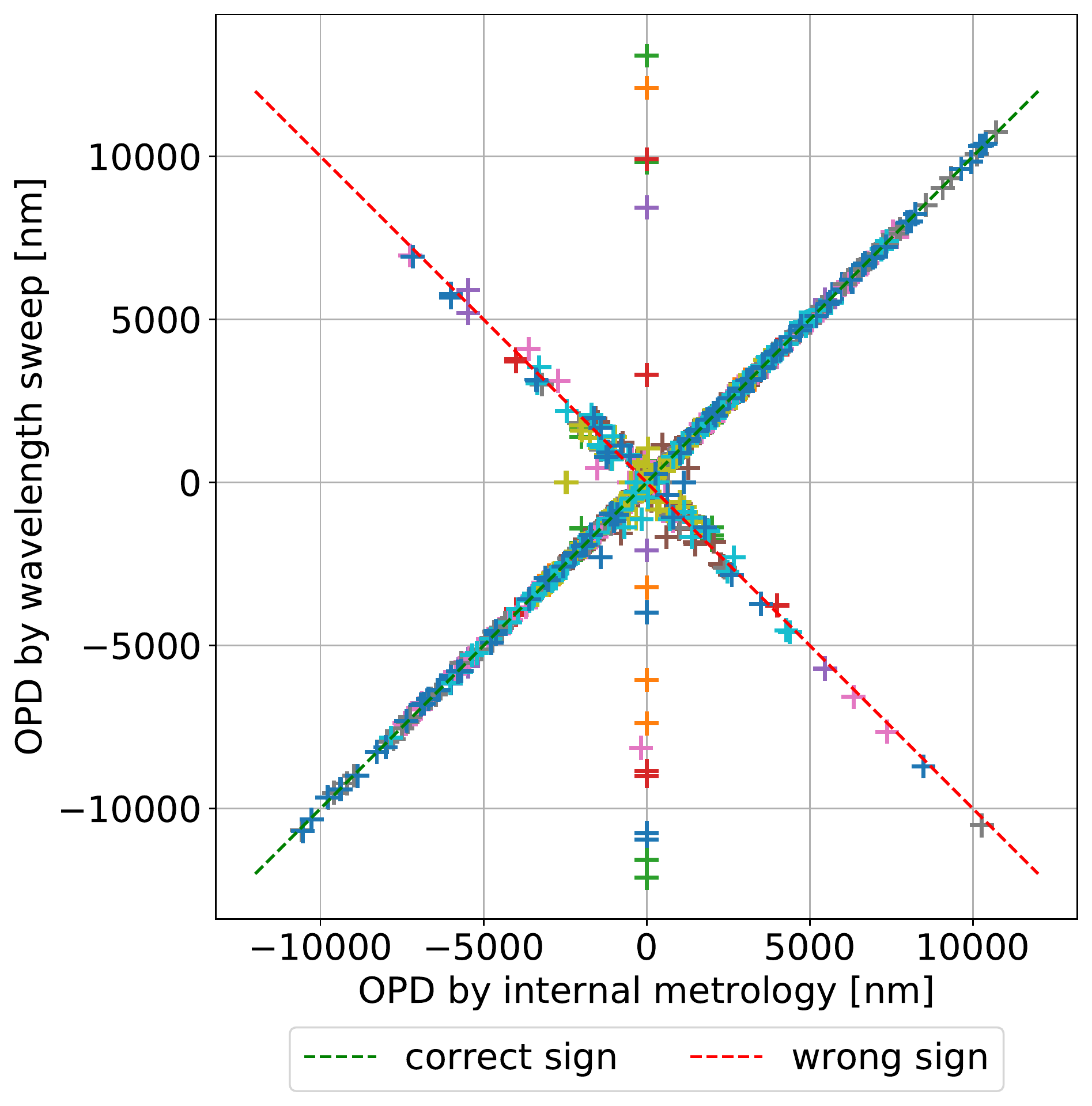}
    \caption{$OPD$ by the wavelength sweep versus imposed $OPD$. The different colors of the crosses denote each configuration that is applied to the  segmented mirror. The crosses show the result of the simulation with an exaggerated alignment error of 0.5 phase mask radius.}
    \label{fig:crossSim05}
\end{figure}

\begin{figure}[h!]
    \centering
    \includegraphics[width = 7cm]{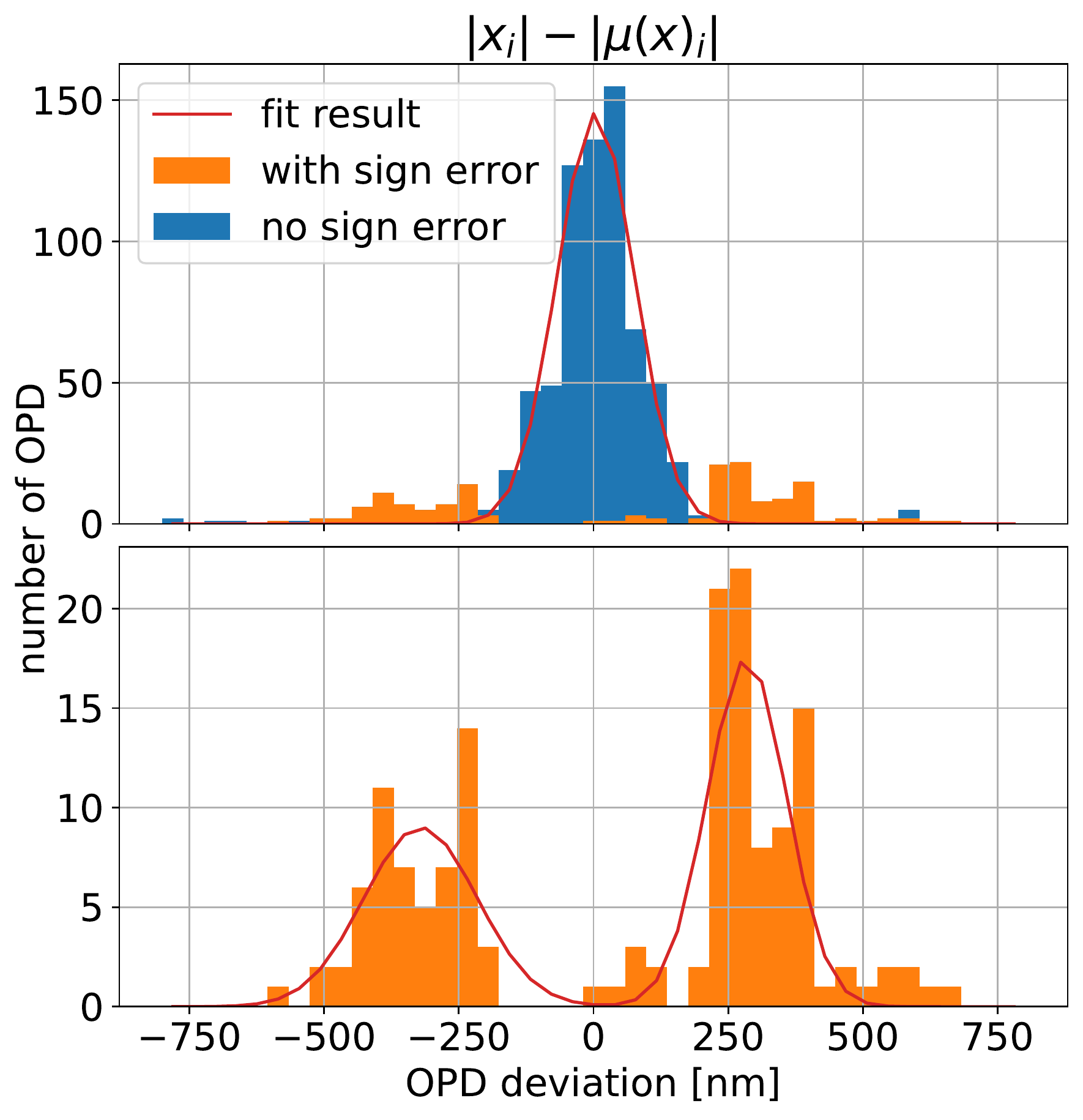}
    \caption{Distribution of the difference between the absolute value of the imposed $OPD$ and the absolute value of reconstructed $OPD$. The top diagram shows the histogram for all data together and the bottom one only for the data with known sign error. The result of both fits are listed in \tabref{tab:fitSim}. In the top diagram with and without sign error are stacked.}
    \label{fig:histSim05}
\end{figure}

\figref{fig:simcross} to \figref{fig:histSim05} summarise these findings. 
\figref{fig:simcross} and \figref{fig:simhist} show the simulated results with the quantity of noise as expected from measurements of the test bench. 

\begin{table}[h!]
\centering
\caption{Gaussian fit parameters to the histograms in \figref{fig:simhist} and \figref{fig:histSim05}}
\begin{tabular}{|m{3cm}|c|c|}
\hline
\multirow{2}{*}{Figure} & \multicolumn{2}{c|}{Fit result} \\ \cline{2-3} 
                  & $\mu$ (nm)&$\sigma$ (nm)\\ \hline
\figref{fig:simhist} top (only without sign error)& 4 & 60 \\ \hline
\multirow{2}{3cm}{\figref{fig:simhist} bottom (only with sign error)} & -338 & 60 \\ \cline{2-3} 
                  & 292 & 13 \\ \hline
 \figref{fig:histSim05} top (only without sign error)& 4 & 71 \\ \hline
\multirow{2}{3cm}{\figref{fig:histSim05} bottom  (only with sign error)} & -320 & 105 \\ \cline{2-3} 
                  & 284 & 73 \\ \hline
\end{tabular}
    \label{tab:fitSim}
\end{table}

First, the fitting of the precision is similar to the one found with the measurement data. 
Second, the sign error shows a similar behaviour, by also showing offsets to both sides of the expected value. 
Only the quantity of sign error is not comparable: only 5\,\% of the borders present sign errors in the simulated data. 
The only way to come closer to the measurement data with 20\,\% of sign errors is to increase the phase mask alignment error to 0.5 x radius. 
This is presented in \figref{fig:crossSim05} and \figref{fig:histSim05}. 

Finally,according to \eqnref{eqn:a1better}, $\phi$ is solely influenced by the misalignment, however, the present simulation disagrees with this statement.
The fit of $\phi$ in the data without noise leads to randomly scattered $\phi$ values inside its boundaries. 
This means that the fit of the phase contrast signal already introduces noise. 
To be exact, the noise comes from the model that incompletely describes the signal. 
\eqnref{eqn:zeus} has been developed from a 1 dimensional model of an isolated border without a gap separating both segments. 
There are features of the signals it approximates, such as the effect of the gap and the slight cross talk between borders, which makes the current fit inexact. 
Developing a more complete model of the phase contrast signal should improve this.

This simulation demonstrates that the technique presented here is  most sensitive to the phase mask alignment. 
The next worse offender is the pupil registration and particularly the orientation of the pupil.
The simulation shows that the drift in $\phi$ is not solely the effect of a misalignment, but also of a fitting noise.
Finally, it also demonstrates that the fit of the phase contrast signal is responsible for noise and should be improved.

\section{Discussion and conclusion}\label{sec:concl}
In this paper we detail a technique dedicated to the accurate measurement of local pistons of segmented mirrors for large telescopes. This technique, named the wavelength sweep, is combined with the principle of the Zernike phase contrast sensor, that has been developed a decade ago in the framework of the VLT-APE experiment. In this paper, we present the simulations of the entire technique from the simulated images to the measurement precision estimation of the piston per segment. This technique has been applied on the dedicated test bench at ESO, to evaluate the performance of the method and compare the results to simulations. 

The wavelength sweep combined with the phase contrast sensor shows promising performances with 64\,nm RMS of precision. When there are no sign errors, a segmented mirror reconstruction can reach a precision as low as 42\,nm RMS using the method. The simulation shows that the technique is resilient to a number of noise sources, including movement of the segments between and during each wavelength acquisition (within a 5\,nm RMS over the entire measurement hence 30 seconds times 16 plus 10 seconds to change the wavelength times 15), a defocus between wavelengths, segment misfiguration, and camera readout noises.The simulation also shows that the measurement technique is more sensitive to alignment of the phase mask.

We demonstrate that the sign error previously investigated \cite{Bonaglia:10}, has its origin in a weakness of the least square criterion to the noise. The sign error is accompanied by a bias that degrades the measurement. A method to correct the sign via the phase closure algorithm is offered, but it does not solve the bias. This degrades the ultimate performance of the final reconstruction from 42\,nm RMS to 112\,nm RMS. However, a higher degree of interconnection between the phase closure algorithm and the wavelength sweep could also improve the sign error correction, allowing removal of the bias. The study in \cite{Pinna:thesis}, suggests that a smaller sampling of the wavelength sweep and several measurements at the same wavelength improve the sign error detection. This remains to be tested with the phase contrast sensor.

We have extensively reviewed the influence of the piston on the measurement. The tip tilt is of concern as it degrades the measurement. The phase contrast sees an angle like a piston variation along the border. In the presented method we average the signal along the borders. The presence of a tilt makes the signal shape change along the border and, if too important, will create a signal shape that is impossible to fit. Ideally, the angle should create less than $\lambda_{s}/10$ of $OPD$ variation, in order to be ignored. This is an angle of 0.025\," on a 90\,cm edge to edge segment. This study assumes that the tip-tilt is dealt with, before the piston, by other means. For example, the tip-tilt could be addressed using the algorithm in \cite{Surdej:10} section 3.D. Instead of a single box, two smaller boxes are used to retrieve the signal along each border and fitted accordingly. Also an interaction matrix including tip tilt is used for the reconstruction. However, this is highly speculative and remains to be tested in future studies. 

The use of the phase contrast sensor in this study is linked to the hardware availability and to the sensor's ease of integration. In terms of performance on sky, \cite{Gonte:11} demonstrate similar performances for the  pyramid and Zernike phase contrast sensors. The strength of the phase contrast sensor lies in its higher tolerance to misalignment and its non-intrusiveness. The pyramid sensor requires a modulation of the star image on the pyramid\cite{pyra:20}. The Zernike phase contrast sensor is less intrusive than the pyramid because, after the phase mask, the beam can  also be used for other applications.

Further improvements to this method can be studied in a forthcoming work, for example the estimation of the impact of the phase contrast signal sampling, the impact of the flux variation vs the wavelength, the optimization of the phase mask parameters (depth and diameter) and the phase contrast signal normalisation.
 
\section*{Disclosure}

The authors declare no conflicts of interest.

\bibliography{sample}

\begin{thebibliography}{10}
\newcommand{\enquote}[1]{``#1''}

\bibitem{ELT:18}
\enquote{The elt in 2017: The year of the primary mirror.} in \emph{The
  Messenger,}  vol. 171 (2018), p. 20–23.

\bibitem{ELT:WS}
{ESO}, \enquote{{The European Extremly Large Telescope "ELT" Project},}
  \url{http://www.eso.org/sci/facilities/eelt/}.

\bibitem{TMT:WS}
{TMT International Observatory}, \enquote{{TMT internationnal Observatory},}
  \url{https://www.tmt.org/}.

\bibitem{GMT:WS}
{GMTO Corporation}, \enquote{{Giant Magelan Telescope},}
  \url{https://www.gmto.org/}.

\bibitem{JWST:WS}
D.~John, \enquote{{James Webb Space Telescope},}
  \url{https://www.jwst.nasa.gov/}.

\bibitem{Keck:91}
B.~{Martin}, J.~M. {Hill}, and R.~{Angel}, \enquote{{The new ground-based
  optical telescopes},} {\protect\JournalTitle{Physics Today}} \textbf{44},
  22--30 (1991).

\bibitem{GTC:04}
P.~{Alvarez} and J.~M. {Rodr{\'\i}guez-Espinosa}, \emph{{The GTC project: in
  the midst of integration}} (2004), vol. 5489 of \emph{Society of
  Photo-Optical Instrumentation Engineers (SPIE) Conference Series}, pp.
  583--591.

\bibitem{GTCPhasing:ukn}
{Grantecan S.A.}, \enquote{{Description of the GTC},}
  \url{http://www.gtc.iac.es/observing/GTCoptics.php}.

\bibitem{Bely:03}
P.~Y. {Bely} and R.-R. {Rohloff}, \enquote{{Book Review: The design and
  construction of large optical telescopes / P. Y. Bely (ed.), Springer,
  Heidelberg, 2003, XXIV+505 pp. ISBN 0-387-95512-7},}
  {\protect\JournalTitle{Sterne und Weltraum}} \textbf{42}, 90 (2003).

\bibitem{LewisSPIE:18}
S.~A.~E. {Lewis}, E.~{Brunetto}, A.~{F{\"o}rster}, C.~{Frank}, I.~{Guidolin},
  S.~{Guisard}, P.~{Hammersley}, R.~{Holzl{\"o}hner}, P.~{Jolley},
  J.~{Kosmalski}, U.~{Lampater}, E.~{Marchetti}, P.~{La Penna}, T.~{Pfrommer},
  and P.~{Zuluaga}, \enquote{{Extremely Large Telescope Prefocal Station A
  system concept},} in \emph{Society of Photo-Optical Instrumentation Engineers
  (SPIE) Conference Series,}  vol. 10700 (2018), p. 107001B.

\bibitem{Chanan:98}
G.~Chanan, M.~Troy, F.~Dekens, S.~Michaels, J.~Nelson, T.~Mast, and D.~Kirkman,
  \enquote{Phasing the mirror segments of the keck telescopes: the
  broadbandphasing algorithm,} {\protect\JournalTitle{Appl. Opt.}} \textbf{37},
  140--155 (1998).

\bibitem{Gonte:11}
F.~Gonte, R.~Mazzoleni, I.~Surdej, and L.~Noethe, \enquote{On-sky performances
  of an optical phasing sensor based on a cylindrical lenslet array for
  segmented telescopes,} {\protect\JournalTitle{Appl. Opt.}} \textbf{50},
  1660--1667 (2011).

\bibitem{APESPIE:08}
F.~{Gonte}, C.~{Araujo}, R.~{Bourtembourg}, R.~{Brast}, F.~{Derie},
  P.~{Duhoux}, C.~{Dupuy}, C.~{Frank}, R.~{Karban}, R.~{Mazzoleni},
  L.~{Noethe}, I.~{Surdej}, N.~{Yaitskova}, R.~{Wilhelm}, B.~{Luong},
  E.~{Pinna}, S.~{Chueca}, and A.~{Vigan}, \enquote{{Active Phasing Experiment:
  preliminary results and prospects},} in \emph{Ground-based and Airborne
  Telescopes II,}  vol. 7012 of \emph{Society of Photo-Optical Instrumentation
  Engineers (SPIE) Conference Series} (2008), p. 70120Z.

\bibitem{Surdej:10}
I.~Surdej, N.~Yaitskova, and F.~Gonte, \enquote{On-sky performance of the
  zernike phase contrast sensor for the phasing of segmented telescopes,}
  {\protect\JournalTitle{Appl. Opt.}} \textbf{49}, 4052--4062 (2010).

\bibitem{Dohlen:03}
K.~{Dohlen}, \enquote{{Phase masks in astronomy: From the Mach-Zehnder
  interferometer to Coronographs},} {\protect\JournalTitle{arXiv e-prints}}
  arXiv:1807.09852 (2018).

\bibitem{VievardSPIE:18}
S.~{Vievard}, F.~{Cassaing}, L.~M. {Mugnier}, A.~{Bonnefois}, and J.~{Montri},
  \enquote{{Real-time full alignment and phasing of multiple-aperture imagers
  using focal-plane sensors on unresolved objects},} in \emph{Society of
  Photo-Optical Instrumentation Engineers (SPIE) Conference Series,}  vol.
  10698 (2018), p. 106986F.

\bibitem{Janin:16}
P.~{Janin-Potiron}, P.~{Martinez}, P.~{Baudoz}, and M.~{Carbillet},
  \enquote{{The self-coherent camera as a focal plane phasing sensor},} in
  \emph{EAS Publications Series,}  vol. 78-79 of \emph{EAS Publications Series}
  (2016), pp. 287--305.

\bibitem{BonnetSPIE:18}
H.~{Bonnet}, F.~{Biancat-Marchet}, M.~{Dimmler}, M.~{Esselborn},
  N.~{Kornweibel}, M.~{Le Louarn}, P.-Y. {Madec}, E.~{Marchetti},
  M.~{M{\"u}ller}, S.~{Oberti}, J.~{Paufique}, L.~{Pettazzi}, B.~{Sedghi},
  J.~{Spyromilio}, S.~{Stroebele}, C.~{V{\'e}rinaud}, and E.~{Vernet},
  \enquote{Adaptive optics at the eso elt,} in \emph{SPIE proceeding,}  vol.
  10703 of \emph{Society of Photo-Optical Instrumentation Engineers (SPIE)
  Conference Series} (2018), p. 1070310.

\bibitem{Chanan:00}
G.~Chanan, C.~Ohara, and M.~Troy, \enquote{Phasing the mirror segments of the
  keck telescopes ii: the narrow-band phasing algorithm,}
  {\protect\JournalTitle{Appl. Opt.}} \textbf{39}, 4706--4714 (2000).

\bibitem{Vigan:11}
A.~Vigan, K.~Dohlen, and S.~Mazzanti, \enquote{On-sky multiwavelength phasing
  of segmented telescopes with the zernike phase contrast sensor,}
  {\protect\JournalTitle{Appl. Opt.}} \textbf{50}, 2708--2718 (2011).

\bibitem{Bonnet:14}
H.~{Bonnet}, M.~{Esselborn}, N.~{Kornweibel}, and P.~{Dierickx}, \enquote{{Fast
  optical re-phasing of segmented primary mirrors},} in \emph{Ground-based and
  Airborne Telescopes V,}  vol. 9145 of \emph{Society of Photo-Optical
  Instrumentation Engineers (SPIE) Conference Series} (2014), p. 91451U.

\bibitem{Bonaglia:10}
M.~{Bonaglia}, E.~{Pinna}, A.~{Puglisi}, S.~{Esposito}, J.~C. {Guerra},
  R.~{Myers}, and N.~{Dipper}, \enquote{{First cophasing of a segmented mirror
  with a tunable filter and the pyramid wavefront sensor},} in \emph{Modern
  Technologies in Space- and Ground-based Telescopes and Instrumentation,}
  vol. 7739 of \emph{Society of Photo-Optical Instrumentation Engineers (SPIE)
  Conference Series} (2010), p. 77392Y.

\bibitem{lctfThor:ukn}
{Thorlabs}, \enquote{{Liquid Crystal Tuneable Filter model KURIOS-XE2/M},}
  \url{https://www.thorlabs.com/thorproduct.cfm?partnumber=KURIOS-XE2/M}.

\bibitem{Gonte:04}
F.~Y.~J. {Gont{\'e}}, N.~{Yaitskova}, P.~{Dierickx}, R.~{Karban},
  A.~{Courteville}, A.~{Schumacher}, N.~{Devaney}, S.~{Esposito}, K.~{Dohlen},
  M.~{Ferrari}, and L.~{Montoya}, \enquote{{APE: a breadboard to evaluate new
  phasing technologies for a future European Giant Optical Telescope},} in
  \emph{Ground-based Telescopes,}  vol. 5489 of \emph{Society of Photo-Optical
  Instrumentation Engineers (SPIE) Conference Series} J.~{Oschmann},
  Jacobus~M., ed. (2004), pp. 1184--1191.

\bibitem{PinnaSPIE:06}
E.~{Pinna}, S.~{Esposito}, A.~{Puglisi}, F.~{Pieralli}, R.~M. {Myers},
  L.~{Busoni}, A.~{Tozzi}, and P.~{Stefanini}, \enquote{{Phase ambiguity
  solution with the Pyramid Phasing Sensor},} in \emph{Society of Photo-Optical
  Instrumentation Engineers (SPIE) Conference Series,}  vol. 6267 (2006), p.
  62672Y.

\bibitem{Michelson:1895}
M.~A. Albert and B.~J. Ren{\'e}, \enquote{D{\'e}termination exp{\'e}rimentale
  de la valeur du m{\`e}tre en longueurs d'ondes lumineuses,}
  {\protect\JournalTitle{Trav. Et Mem. Bur. Int. Poids es Mes.}} \textbf{11},
  1--42 (1895).

\bibitem{Pinna:thesis}
E.~Pinna, \enquote{Study and characterization of the pyramid wavefront sensor
  for co-phasing.} Ph.D. thesis, UNIVERSITA DEGLI STUDI DI FIRENZE (2010).

\bibitem{ObsAstr:12}
P.~{L{\'e}na}, D.~{Rouan}, F.~{Lebrun}, F.~{Mignard}, and D.~{Pelat},
  \emph{{Observational Astrophysics}} (2012). Section 6.4.6 page 271-274.

\bibitem{hecht:book}
E.~{Hecht}, \emph{Optics} (Addison-Wesley, 1998), chap. 13.2.4, pp. p611--615,
  3rd ed.

\bibitem{Kolb:06}
J.~{Kolb}, S.~{Oberti}, E.~{Marchetti}, and F.~{Quir{\'o}s-Pacheco},
  \enquote{{Full characterization of the turbulence generator MAPS for MCAO},}
  (2006), p. 627258.

\bibitem{Wilhelm:08}
R.~Wilhelm, B.~Luong, A.~Courteville, S.~Estival, F.~Gont\'{e}, and
  N.~Schuhler, \enquote{Dual-wavelength low-coherence instantaneous
  phase-shifting interferometer to measure the shape of a segmented mirror with
  subnanometer precision,} {\protect\JournalTitle{Appl. Opt.}} \textbf{47},
  5473--5491 (2008).

\bibitem{Surdej:thesis}
I.~Surdej, \enquote{Co-phasing with phase the phase contrast sensor,} Ph.D.
  thesis, Universitee de Liege (2010).

\bibitem{Surdej:07}
I.~{Surdej}, H.~{Lorch}, L.~{Noethe}, N.~{Yaitskova}, and R.~{Karban},
  \emph{{Pattern recognition and signal analysis in a Mach-Zehnder type phasing
  sensor}} (2007), vol. 6696 of \emph{Society of Photo-Optical Instrumentation
  Engineers (SPIE) Conference Series}, p. 66960L.

\bibitem{pyra:20}
I.~Shatokhina, V.~Hutterer, and R.~Ramlau, \enquote{{Review on methods for
  wavefront reconstruction from pyramid wavefront sensor data},}
  {\protect\JournalTitle{Journal of Astronomical Telescopes, Instruments, and
  Systems}} \textbf{6}, 1 -- 39 (2020).

\end{thebibliography}






\end{document}